\documentclass[aps,a4paper,showpacs,preprintnumbers,amsmath,amssymb,floatfix]{revtex4}
\usepackage{dcolumn}
\usepackage{bm}
\usepackage{amsmath,amssymb}
\usepackage{booktabs, subfigure}
\usepackage{graphicx,psfrag}
\usepackage{psfrag}
\usepackage{epsfig}
\usepackage{epstopdf}
\usepackage{color} 
\usepackage{rotating}
\usepackage{slashed}
\usepackage{geometry}
\usepackage{rotating}
\usepackage{tikz}
\usetikzlibrary{shapes.geometric, arrows}
\allowdisplaybreaks[3]
\newcommand{\M}{\ensuremath{\mathcal{M}}}

\numberwithin{equation}{section} 
\def\lsim{\raise0.3ex\hbox{$\;<$\kern-0.75em\raise-1.1ex\hbox{$\sim\;$}}}
\def\gsim{\raise0.3ex\hbox{$\;>$\kern-0.75em\raise-1.1ex\hbox{$\sim\;$}}}
\newcommand{\beq}{\begin{equation}}
\newcommand{\eeq}{\end{equation}}
\newcommand{\bal}{\begin{align}}
\newcommand{\eal}{\end{align}}
\newcommand{\bc}{\begin{cases}}
\newcommand{\ec}{\end{cases}}
\newcommand{\br}{\begin{array}}
\newcommand{\er}{\end{array}}
\newcommand{\rf}[1]{(\ref{#1})}

\newcommand{\mgr}{m_{3/2}}
\newcommand{\dya}{D_{Y\alpha}}
\newcommand{\db}{D_{B_{13}}}
\newcommand{\dx}{D_{\chi_1}}

\newcommand{\dy}{D_{Y_{13H}}}
\newcommand{\dz}{D_Z}
\newcommand{\muplus}{\mu_{\text{MESS}}^+}
\newcommand{\muminus}{\mu_{\text{MESS}}^-}
\newcommand{\mugut}{\mu_{\text{GUT}}}
\newcommand{\mutev}{\mu_{\text{TEV}}}
\newcommand{\mumes}{\mu_{\text{MESS}}}
\newcommand{\tmes}{t_{\text{MESS}}}
\newcommand{\iba}{I_{B_1}}
\newcommand{\ima}{I_{M_1}}
\newcommand{\ibb}{I_{B_2}}

\newcommand{\ibc}{I_{B_3}}
\newcommand{\imc}{I_{M_3}}
\newcommand{\ib}[1]{I_{B_{#1}}}
\newcommand{\im}[1]{I_{M_{#1}}}
\newcommand{\nalpha}{n_{\alpha}}
\newcommand{\nbeta}{n_{\beta}}
\newcommand{\ngamma}{n_{\gamma}}
\newcommand{\ndelta}{n_{\delta}}
\newcommand{\nepsilon}[1]{n_{\epsilon_{#1}}}
\newcommand{\ndg}{n_{\delta\gamma}}
\newcommand{\gGUT}{g_{\text{GUT}}}
\newcommand{\rv}{\nonumber\\&}
\newcommand{\gmes}[2]{g_{#1}^#2(\mumes)}
\newcommand{\ibcal}[1]{\mathcal{I}_{B_{#1}}}
\newcommand{\imcal}[1]{\mathcal{I}_{M_{#1}}}
\newcommand{\Ycal}[1]{\mathcal{Y}(#1)}
\newcommand{\ag}{\alpha_g}

\begin{document}
\begin{flushright}
HIP-2014-38/TH
\end{flushright}
\title{ Renormalization group invariants and sum rules in the deflected 
mirage mediation supersymmetry breaking}
\author{Katri Huitu, $^1$
\footnote{\tt{Electronic address: katri.huitu@helsinki.fi}}
P.  N. Pandita, $^2$
\footnote{\tt{Electronic address: pandita@iucaa.ernet.in}} and
Paavo Tiitola $^1$
\footnote{\tt{Electronic address: paavo.tiitola@helsinki.fi }}}
\affiliation{ $^1$ Department of Physics, and Helsinki Institute of
Physics,
P. O. Box 64, FIN-00014 University of Helsinki, Finland}
\affiliation{ $^2$ Centre of High Energy Physics, Indian Institute of Science, 
Bangalore 560 012, India}
\thispagestyle{myheadings}

\begin{abstract}
\noindent
We examine the deflected mirage mediation supersymmetry breaking (DMMSB) scenario, which combines three supersymmetry breaking scenarios, namely anomaly mediation, gravity mediation and gauge mediation using the one-loop renormalization group invariants (RGIs). We examine the effects on the RGIs at the threshold where the gauge messengers emerge, and derive the supersymmetry breaking parameters in terms of the RGIs. 
We further discuss whether  supersymmetry breaking mediation mechanism can be determined using a limited set of invariants, and derive sum rules valid for DMMSB below the gauge messenger scale.
In addition we examine the implications of the measured Higgs mass for the DMMSB spectrum.

\end{abstract}
\pacs{11.30.Pb, 12.60.Jv, 14.80.Ly}
\maketitle


\section{Introduction}
\label{sec:Introduction}
Supersymmetry remains at present a favored solution to the hierarchy 
problem of the standard model~(SM).
Supersymmetry is obviously a broken 
symmetry, as none of the superpartners of the SM particles have been
experimentally observed so far. 
The idea of weak scale supersymmetry,
as embodied, for example,  in the minimal supersymmetric standard model~(MSSM), 
as a solution to the gauge hierarchy problem 
has already been partly tested at the Large Hadron Collider~(LHC) \cite{ATLASsusy, CMSsusy}.
The most minimal constrained model is found to be disfavoured, and
the emphasis in the studies has moved to other well motivated models.

The mechanism of supersymmetry
breaking, which is crucial for determining the masses of the superpartners of the 
SM particles,  is not known at present. There are  different viable 
models of supersymmetry breaking in which supersymmetry  breaking 
is mediated  by a specific interaction in a hidden sector and is communicated 
to the visible sector via some mediator fields. 
Well known examples  of hidden sector supersymmetry breaking  include
gravity mediated supersymmetry breaking~\cite{Chamseddine:1982jx}, gauge mediated supersymmetry 
breaking~\cite{Dine:1994vc},
and anomaly mediated supersymmetry breaking~\cite{Randall:1998uk}, respectively. 
It is usually assumed that one mediation mechanism dominates, which depends
on the type of problem that one wishes to address in the context of
MSSM. 
In fact, it may well be that a single mediation mechanism does not dominate. 
In some 
situations the anomaly mediated and gravity mediated contributions to the 
supersymmetry breaking can coexist and manifest at a comparable
value, leading to what is known as mirage mediation, 
a situation that naturally arises in some string theories~\cite{Kachru:2003aw}.
The name derives from the 
fact that soft gaugino masses unify at a scale~(the mirage scale) 
which is lower than the grand unification or GUT scale. Models based on mirage mediation
can solve the tachyonic slepton mass problem of the 
anomaly mediation models. The mirage mediation models have been
studied extensively in the literature~\cite{Choi:2004sx, Choi:2005hd,
Choi:2005uz, Choi:2007ka}.

However, in the absence of any definite experimental indication
regarding the mass spectrum of the sparticles, 
it is important to consider the case where all the three 
types of supersymmetry breakings contribute to the
soft masses, and ultimately determine the mass spectrum of 
the supersymmetric partners of the SM states. Such a scenario
has been dubbed as  deflected mirage mediation~\cite{hep-ph/0504037}.
In the deflected mirage mediation
the gaugino mass unification is deflected by the threshold effect from 
the messenger fields associated with the gauge mediation contribution.
The messenger fields are included at scales $\mu>\mu_{\rm mess}$, and the 
renormalization group running is affected by the threshold corrections at
this scale.  It is worthwhile to note that in a broad class of 
supergravity models, which might be realized in string theory, the 
contributions to soft supersymmetry breaking masses from 
the three different mediation mechanisms are comparable \cite{Everett:2008ey}.
This makes the study of deflected mirage mediation models rather compelling.


As discussed above, in  order to predict the masses of the superpartners of the SM particles,
it is essential to understand  the nature of supersymmetry breaking.
The usual approach to this has been to consider a particular model
of supersymmetry breaking, e.g. the gravity mediated supersymmetry 
breaking model~(usualy referred to as  mSUGRA) with a limited number
of parameters at the high 
scale~(possibly the GUT scale), 
and evolve these, using renormalization  group equations, 
to the electroweak scale and fit them to the data~\cite{Chattopadhyay:2001va,
Chattopadhyay:2003xi, Djouadi:2006be, Baltz:2006fm, Allanach:2007qk,
Berger:2008cq, Lafaye:2004cn}. This method, the so called ``top-down''
approach has some drawbacks, namely that the reliability of the fit
decreases with the increase of the number of parameters at the high scale.
This approach is also sensitive to uncertainities of the measured
quantities, which includes gauge couplings at the electroweak scale
and the masses of the SM particles.

An alternative approach, which is complementary to the ``top-down''
approach, that has been advocated involves the measured masses at the
elctroweak scale and evolve these to the high scale where supersymmetry is 
broken~\cite{Carena:1996km, Kneur:1998gy, Blair:2002pg, Blair:2005ui}. 
The  resulting structure is then analyzed and conclusions
about the underlying theory at high scales obtained. This approach,
which can be called ``bottom-up'' approach, 
has uncertainities resulting from the present experimental 
uncertainities in the measurement of gauge and Yukawa couplings.

Another  alternative to obtain information on the nature of 
supersymmetry breaking is to obtain specific relations 
among the masses of the superpartners of the SM particles which 
result from the structure of the underlying theory at the high scale.
In this approach the Yukawa couplings of the first- and second-generations
are usually ignored and the renormalization group is used to evolve the 
parameters from high scale to the low scale, and then specific relations 
among sparticle masses are derived  based on the  particular theory at 
the high scale~\cite{Huitu:2002fg, Ananthanarayan:2003ca, 
Ananthanarayan:2004dm, Huitu:2005wh, Ananthanarayan:2007fj, Huitu:2010me,
Huitu:2011yx, Miller:2012vn}. In this context it has also been pointed out that 
there are a set of combinations of parameters of a supersymmetric 
model that are renormalization group invariant~(RGI)  
at the 1-loop level~\cite{Demir:2004aq, Meade:2008wd, Carena:2010gr,
Carena:2010wv, Carena:2012he, Hetzel:2012bk, Hetzel:2012ce}.
Although the argument regarding RGI holds only at the leading-log order, 2-loop corrections
are expected to be small and are likely to be negligible compared with 
the experimental uncertainities. Thus, these higher order effects can either
be neglected or absorbed into a shift of the measured values of the
renormalization group invariants.
It has been argued  that these RGIs can be used to extract the parameters of
different supersymmetry breaking models at the high scale, and thereby
establish the mechanism of supersymmetry breaking. RGIs can also be used for constructing sum rules relating the particle masses, and they are discussed in this context in \cite{Ananthanarayan:2003ca,Ananthanarayan:2004dm,Ananthanarayan:2007fj,Hetzel:2012bk,Hetzel:2012ce,Martin:1993ft}.\par
In this paper we derive the renormalization group invariants for deflected mirage mediation and examine how the appearance of messenger particles at the specific energy
scale lower than the GUT-scale affects the invariants. 
We investigate the GUT-scale parameters of the deflected mirage mediation in terms of the RGIs, and examine validity of sum rules derived previously of the RGIs and derive new ones specific
to the deflected mirage mediation.
In particular, we derive soft supersymmetry breaking parameters in terms of the RGIs.
We also examine the Higgs mass in the deflected mirage mediation scenario to constrain the parameters of the model.

The plan of this paper is as follows. In Section~\ref{sec:DMM}, we describe
the deflected mirage mediation mechanism that we consider in this paper.
Here we analyze the soft supersymmetry breaking gaugino and scalar masses of 
the model, and the boundary conditions on these parameters.
Phenomenological implications, which include the Higgs mass, are studied
in this Section. In Section~\ref{sec:rginvariants} we describe the 
renormalization group invariants and calculate these invariants for the 
deflected mirage mediation supersymmetry breaking model. We then use these
invariants to solve for the parameters of the model.
In Section~\ref{sec:sumrules} we obtain sum rules for the deflected mirage
mediation scenario and discuss some special cases. In 
Section~V we proceed to compare the predictions of the
pure mirage mediation scenario with those of the gravity, AMSB, and the 
gauge mediation of supersymmetry breaking models, respectively. Finally,
in Section~VI we present our conclusions. 

\section{Deflected mirage mediation}
\label{sec:DMM}
The deflected mirage mediation mechanism for supersymmetry 
breaking mechanism  involves contributions of comparable scale from 
gravity mediation, anomaly mediation and gauge mediation  as opposed to the mirage mediation which excludes the contribution from gauge mediation
The quantity 
\begin{equation}
\alpha_m=m_{3/2}/(M_0\log{M_P/m_{3/2}}),
\label{alpham}
\end{equation}
parametrises the anomaly to 
gravity mediation ratio, while $M_0$ describes the mass scale of soft supersymmetry breaking terms \cite{Everett:2008qy}.
Here $m_{3/2}$ is the gravitino mass and $M_P$ the reduced Planck mass.
The ratio of the gauge mediated contribution to its anomaly mediated counterpart
is parametrised by $\alpha_g$. It is related to the messenger fields by the equation
\begin{equation}
\left|\alpha_g\right|=\Lambda/\mgr.
\label{alphag}
\end{equation}
where $\Lambda$ is a mass scale associated with the messenger fields. The absolute value is allows $\alpha_g$ to have negative values.
The parameters  $\alpha_m$ and $\alpha_g$ can be considered continuous but in string 
motivated scenarios they usually have discrete values of the order one~\cite{Everett:2008ey}.
The messenger sector is assumed to come in complete GUT representations in order to 
preserve gauge coupling unification.
 $N$ represents the number of copies of $\mathbf{5},\bar{\mathbf{5}}$ representations under $SU(5)$.
The original Kaluza-Klein compactification is obtained with $\alpha_m$=1 and $N=$0.
Both positive and negative values are possible for $\alpha_g$.  Phenomenological implications of various values of the parameters are discussed in e.g.~\cite{Everett:2008ey} and~\cite{Abe:2014kla}, especially regarding to the Higgs mass.

Above the messenger scale the renormalization group equations are modified from the MSSM 
form by adding the number of messenger pairs to the 
$\beta$-function coefficients $b_a$ \cite{Everett:2008qy}. 
Thus, 
\begin{equation}
b_a'=b_a+N,
\label{bdot}
\end{equation} 
where $\left\{b_1,b_2,b_3\right\}=\left\{33/5,1,-3\right\}$.
At the GUT scale $\mu_{\rm GUT}$, the gaugino mass boundary conditions can be written as \cite{hep-ph/0504037}:
\begin{align}
M_a\left(\mu_\text{GUT}\right) =& M_0\left(1+g_a\ln(M_P/m_{3/2})b_a g_a\alpha_m\right)\nonumber\\ 
					        =& M_0 + g_a^2\left(\mu_\text{GUT}\right)\frac{b_a'}{16\pi^2}m_{3/2},
\qquad (a=1,2,3).
\label{gauginoUV}
\end{align}
Here $\mu_{GUT}$ is the high scale which we take to be the GUT scale. 
Similarly the scalar masses can be written as
\begin{align}
m_i^2(\mu_{GUT})  =&  M^2_0\left[(1-n_i) - \frac{\theta_i}{16\pi^2}\alpha_m\ln (M_{P}/m_{3/2}) 
- \frac{\dot{\gamma'_i}}{(16\pi^2)^2}(\alpha_m\ln(M_P/m_{3/2}))^2\right]\nonumber\\
=&M^2_0(1-n_i) - \frac{\theta_i}{16\pi^2}\mgr M_0- \frac{\dot{\gamma'_i}}{(16\pi^2)^2}\mgr^2,
\label{scalarUV}
\end{align}
where $n_i$ are the modular weights for the scalar masses, $\gamma_i$ are the anomalous dimensions, 
\begin{eqnarray}
\gamma_i = 2 \sum_a g_a^2 c_a(\Phi_i) - \frac{1}{2}\sum_{lm} |y_{ilm}|^2,
\label{gammaexp}
\end{eqnarray}
in which $c_a$ is the quadratic Casimir operator for the field $\Psi_i$, and $y_{ilm}$ are the normalized Yukawa couplings. $\dot{\gamma}$, and $\theta_i$ are defined as
\begin{align}
\dot{\gamma}_i=&2\sum_a g_a^4b_a c_a(\Phi_i) - \sum_{lm} |y_{ilm}|^2b_{y_{ilm}},\\
\theta_i =& 4 \sum_a g_a^2 c_a(\Psi_i) - \sum_{i,j,k} |y_{ijk}|^2
( p- n_i -n_j- n_k),
\end{align}
in which $b_{y_{ilm}}$ is the beta function for the Yukawa coupling $y_{ilm}$. 
$\dot\gamma^\prime_i$ is obtained by replacing $b_a$ with $b'_a = b_a + N$.
For explicit values of  $\theta'_i$, $\dot{\gamma'_i}$ see \cite{Everett:2008qy}. 
One-loop renormalization group equations give the boundary condition at the 
messenger scale $\mu_{\rm mess}$ for the soft gaugino mass parameters:

\begin{equation}
M_a = g_a^2\frac{b_a'}{16\pi^2}m_{3/2} + M_0\left[1-g_a^2\frac{b_a'}{8\pi^2}\log\left(\frac{\mu_\text{GUT}}{\mu_\text{mess}}\right)\right] + \Delta M_a \qquad (a=1,2,3),
\end{equation}
where
\begin{equation}
\Delta M_a = -N M_0\frac{g_a^2}{16\pi^2}\alpha_m(1+\alpha_g) \ln \frac{M_P}{m_{3/2}}
\label{deltaM}
\end{equation}
is a threshold contribution that arises when the messenger fields are integrated out. 
Similarly, scalar masses receive a threshold correction,
\begin{equation}
\Delta {m^{2}_{i}}^{j}=M_0^2\sum_a 2 c_a(\Psi_i) N\frac{g_a^4 (\mu_{\rm mess}) }{(16\pi^2)^2} \left [\alpha_m  (1+\alpha_g)  \ln \frac{M_P}{m_{3/2}}\right ]^2\delta_{i}^{j}.
\label{deltam}
\end{equation}
The gaugino masses unify at the mirage scale \cite{Everett:2008qy}
\begin{equation}
\mu_{\rm mirage} = \mu_{GUT}\left (\frac{m_{3/2}}{M_P} \right )^{\alpha_{\rm m}\rho/2},
\end{equation}
in which 
\begin{equation}
\rho= \frac{1+ \frac{2Ng_0^2}{16\pi^2}  \ln \frac{M_{\rm GUT}}{\mu_{\rm mess}}}{1- \frac{\alpha_{\rm m} \alpha_{\rm g} Ng_0^2}{16\pi^2} 
 \ln \frac{M_P}{m_{3/2}}}.
 \label{miragescale}
 \end{equation}
When $\rho=1$, this reduces to the mirage scale of pure mirage mediation as the deflection is removed. We note that even if gauge mediation is turned
off by setting $\alpha_g=0$, mirage mediation is not recovered. This is  achieved only 
by removing the messenger fields by setting $N=0$. 
This is due to the messenger particles affecting the beta functions and thus anomaly mediation at high scales. 

\begin{figure}
\psfrag{mh0}{$m_{h_0}$(GeV)}
\psfrag{alpha_g}{$\alpha_g$}
  \subfigure[$\alpha_m=$ 0.5, $M_0=$ 3 TeV]{\includegraphics[height=4.9cm]{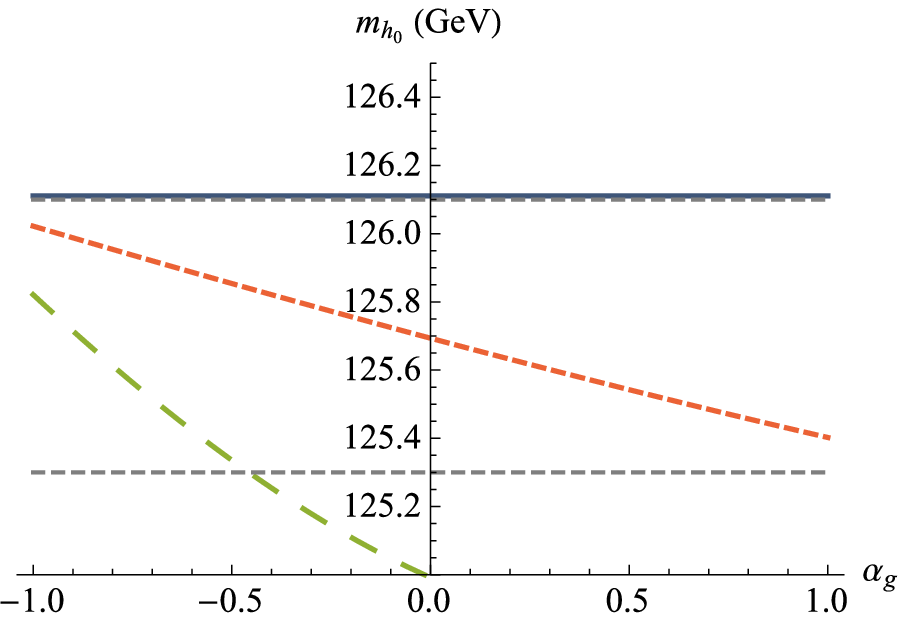}}
  \subfigure[$\alpha_m=$ 1, $M_0=$ 3 TeV]{\includegraphics[height=4.9cm]{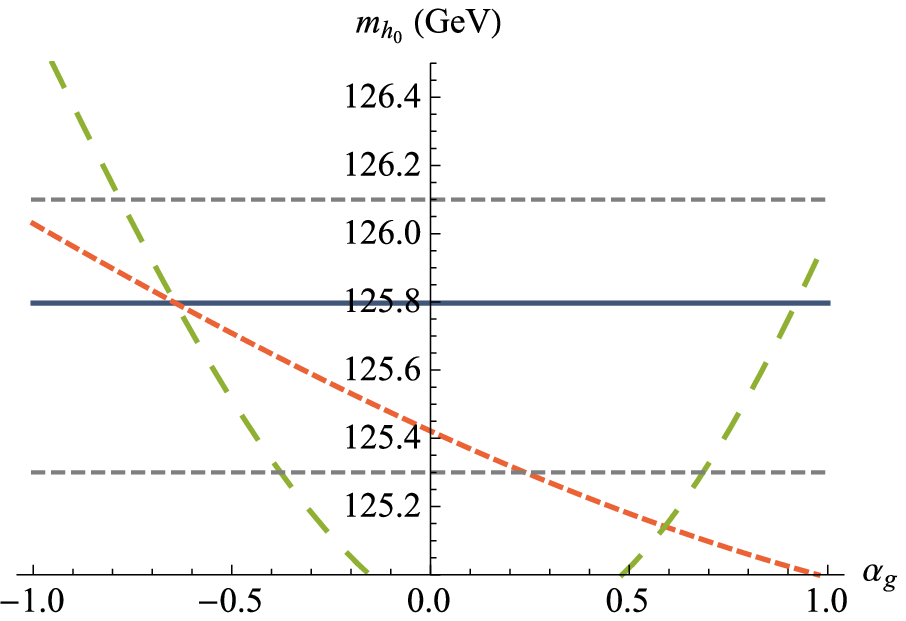}}
  \subfigure[$\alpha_m=$ 2, $M_0=$ 3 TeV]{\includegraphics[height=4.9cm]{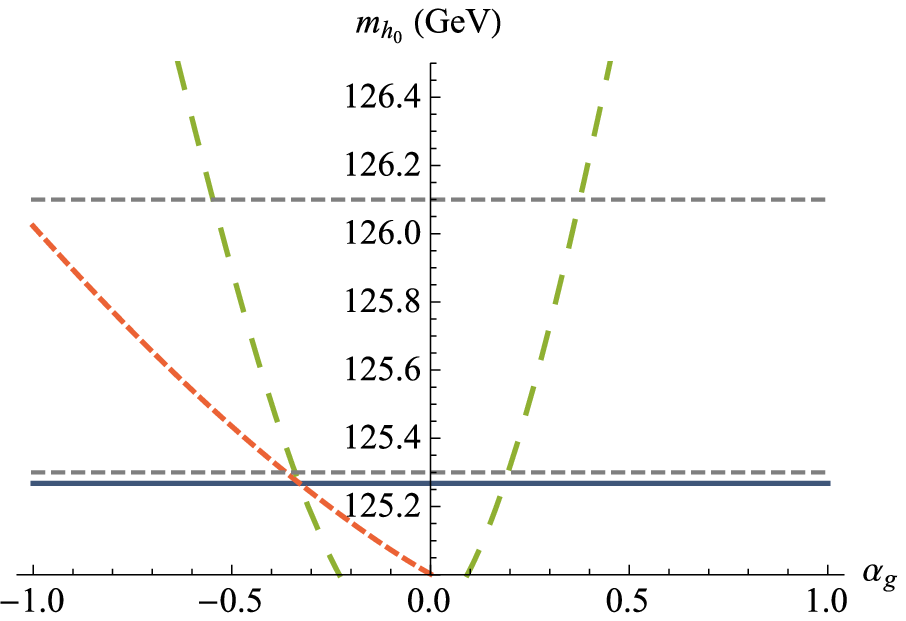}} 
  \subfigure[$\alpha_m=$ 0.5, $M_0=$ 2 TeV]{\includegraphics[height=4.9cm]{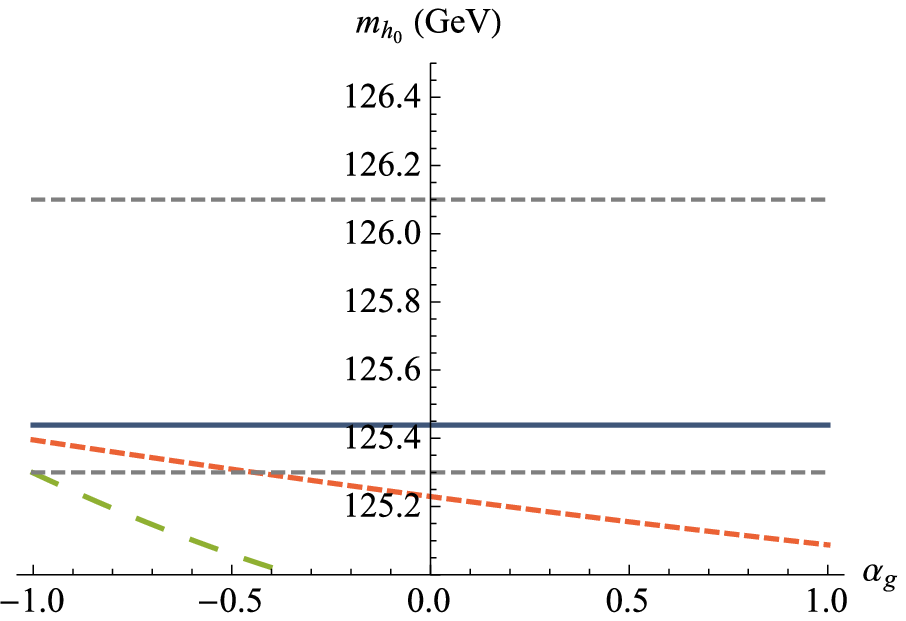}}
  \subfigure[$\alpha_m=$ 1, $M_0=$ 2 TeV]{\includegraphics[height=4.9cm]{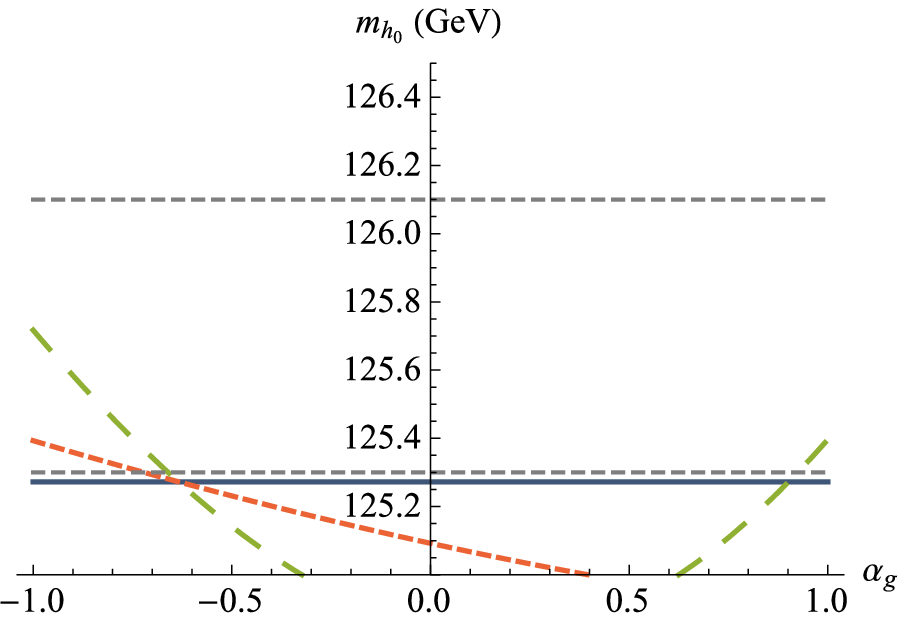}}
  \subfigure[$\alpha_m=$ 2, $M_0=$ 2 TeV]{\includegraphics[height=4.9cm]{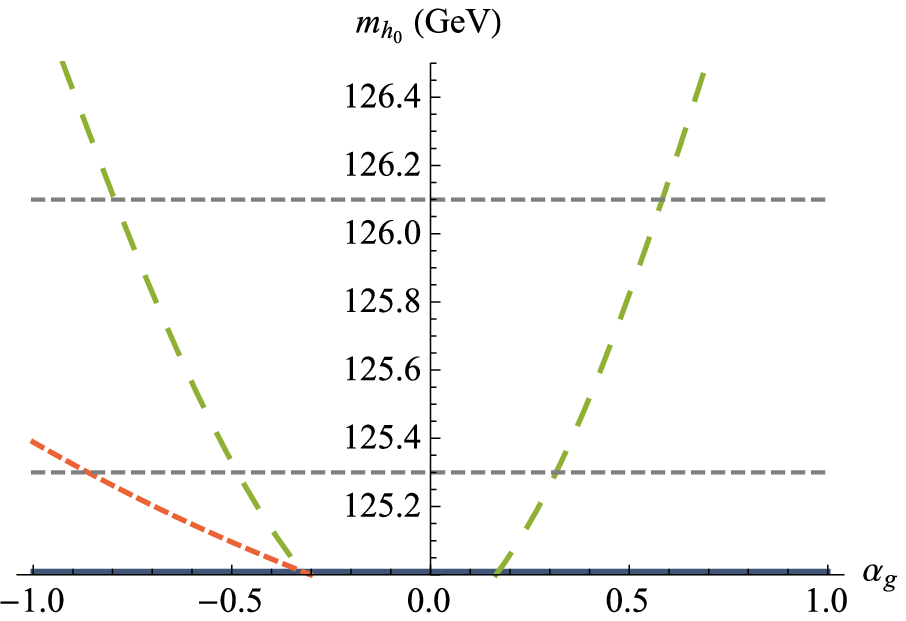}}
\caption{Higgs mass as a function of $\alpha_g$. $n_Q=n_U = n_D = n_L = n_E = $1/2,  and $\mu_{\rm mess}=10^{12}$ GeV. Thick lines in the order of increasing dash length correspond to N= 0 (solid), 3, 10. The Higgs mass is within current experimental limits between the dashed horizontal lines.}
\label{mhiggs}
\end{figure}

To examine the sparticle spectrum of the deflected mirage mediation we have calculated the soft scalar and gaugino masses using one-loop renormalization group equations and the lightest Higgs mass using the one-loop radiative corrections presented in \cite{hep-ph/9504316}. In Fig.~\ref{mhiggs} we have plotted the lightest Higgs mass for several values for the parameters $\alpha_m$, $M_0$ and $N$ as a function of $\alpha_g$. The current experimental limits for the Higgs mass are represented as horizontal dashed lines \cite{Agashe:2014kda}. 
Although not all the parameter combinations satisfy the experimental bounds on the Higgs mass, for any two values of the studied parameters $\alpha_g$, $\alpha_m$, $M_0$, $N$, a viable set can be found.
We note that larger values for $M_0$ and negative values for $\alpha_g$ are favored. Smaller $M_0$ requires small $\alpha_g$ in order to have the Higgs mass in the allowed range.  

\begin{figure}
\psfrag{M}{$m_{\tilde{i}}^2$(GeV)}
\psfrag{G}{$\alpha_g$}
  \subfigure[1. generation. $\alpha_m=$ 0.5, $M_0=$ 3 TeV]{ \includegraphics[height=4.9cm]{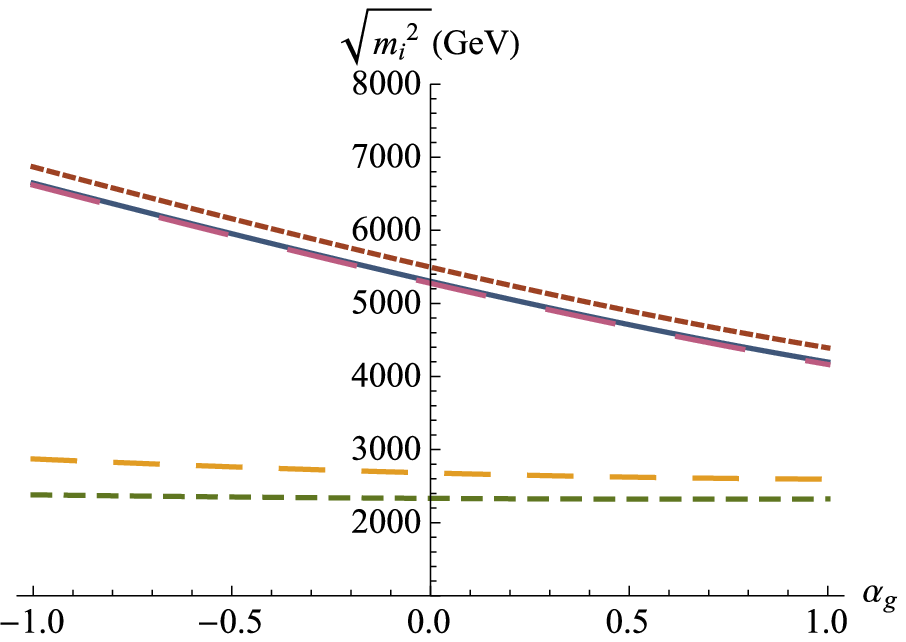}}
  \subfigure[3. generation. $\alpha_m=$ 0.5, $M_0=$ 3 TeV]{\includegraphics[height=4.9cm]{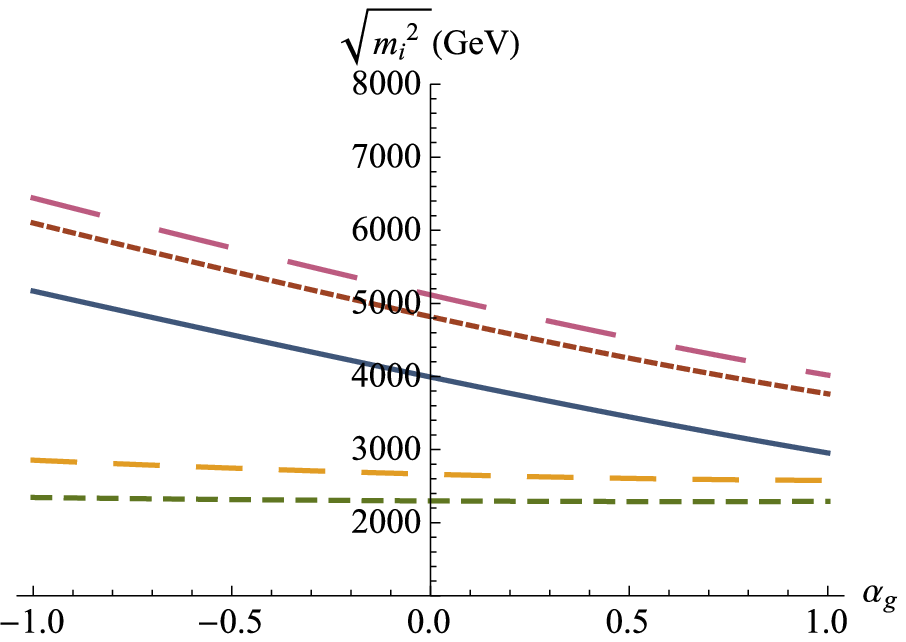}}
  \subfigure[1. generation. $\alpha_m=$ 1.0, $M_0=$ 2 TeV]{\includegraphics[height=4.9cm]{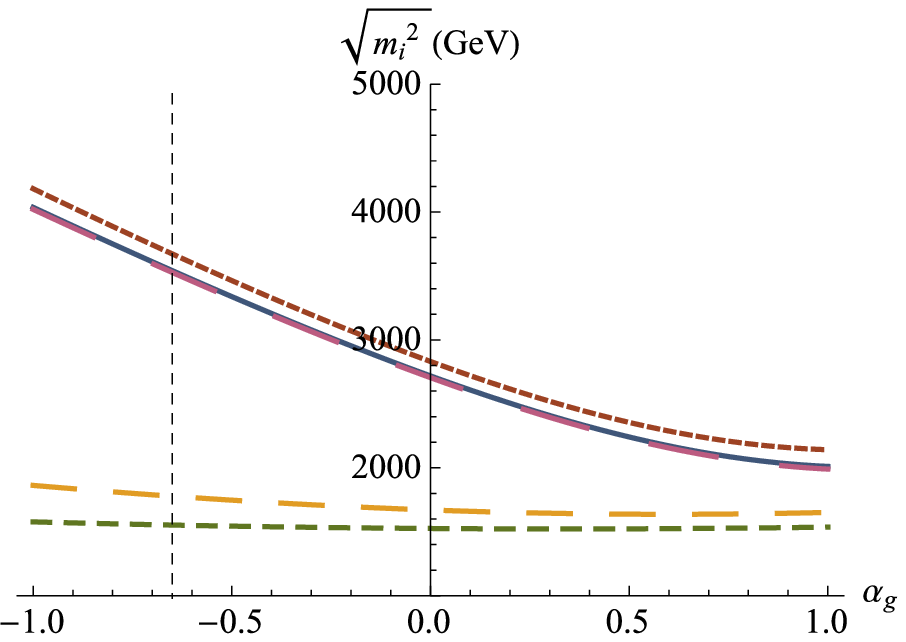}} 
  \subfigure[3. generation. $\alpha_m=$ 1.0, $M_0=$ 2 TeV]{ \includegraphics[height=4.9cm]{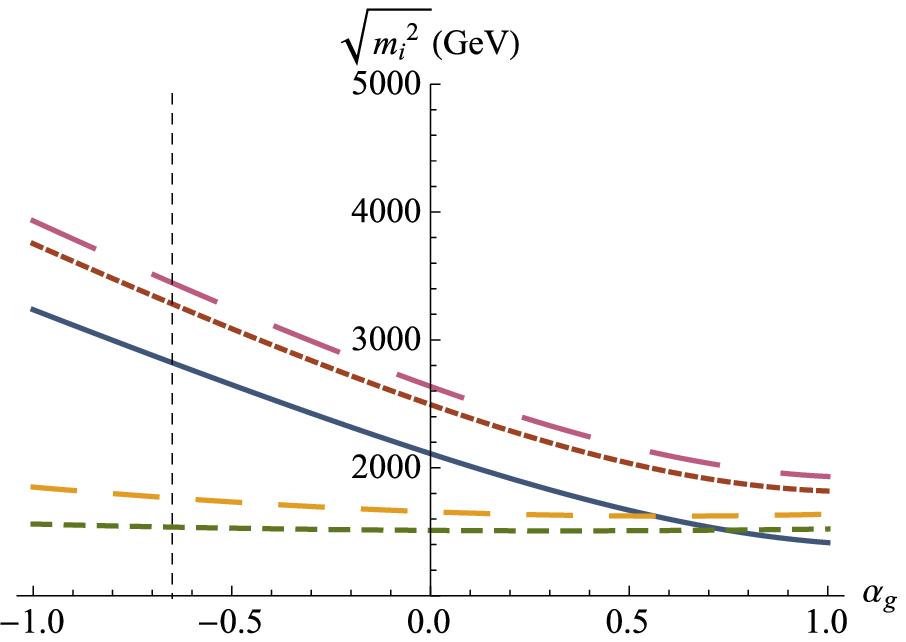}}
\caption{The first and the third generation soft scalar mass parameters plotted as a function of $\alpha_g$. In the order of increasing dash length: $m_{\tilde{U}}$(solid),  $m_{\tilde{Q}_1}$,$m_{\tilde{e}}$,$m_{\tilde{L}}$, and $m_{\tilde{d}}$. $n_Q=n_U = n_D = n_L = n_E = $1/2, $\mu_{\rm mess}=10^{12}$ GeV, and N=3. In (c) and (d) Higgs mass is in the allowed range on the left hand side of the dashed vertical line.}
\label{mscalar}
\end{figure}

\begin{figure}
\psfrag{M}{$M_i$(GeV)}
\psfrag{G}{$\alpha_g$}
\subfigure[$\alpha_m=$ 0.5, $M_0=$ 3 TeV]{ \includegraphics[height=4.9cm]{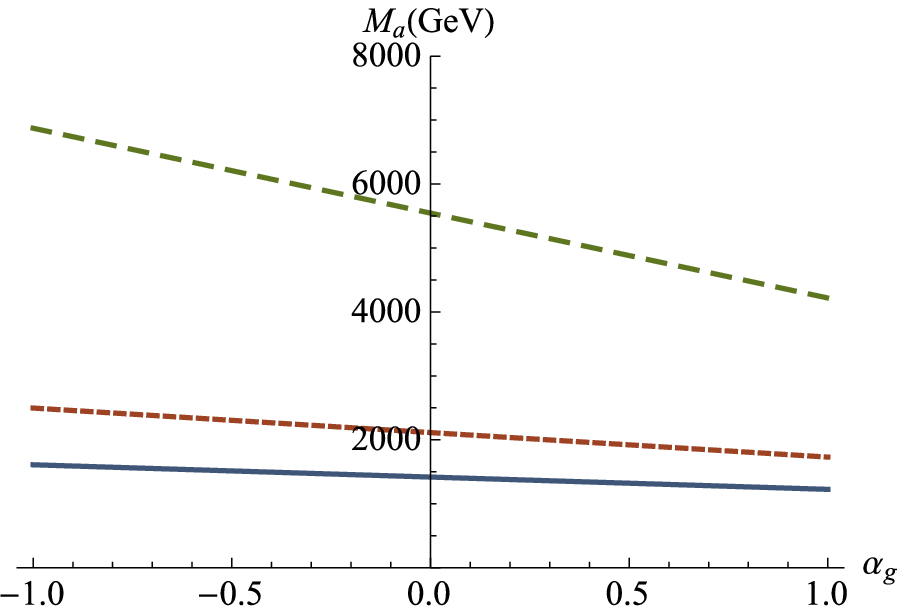}}
 \subfigure[$\alpha_m=$ 1.0, $M_0=$ 2 TeV]{ \includegraphics[height=4.9cm]{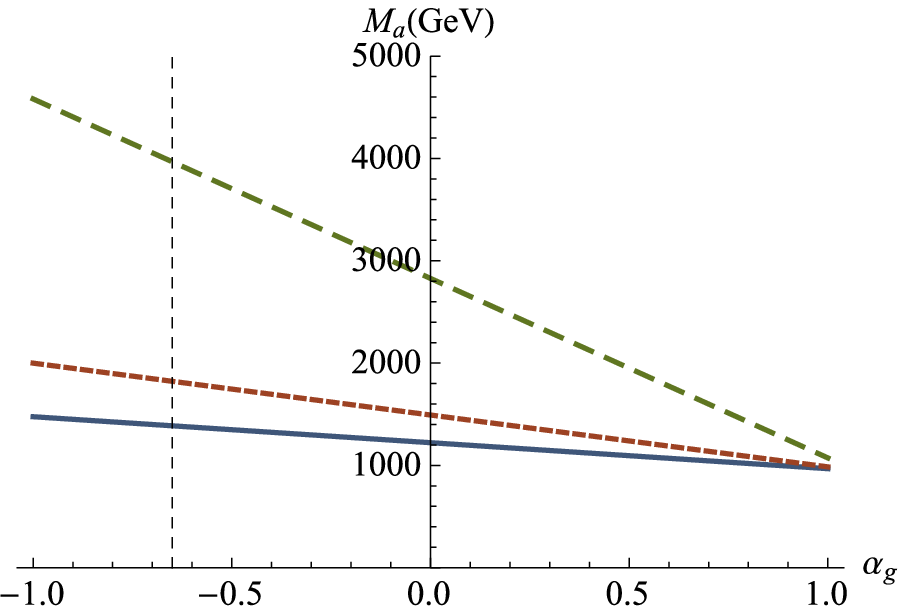}}

\caption{The gaugino masses plotted as a function of $\alpha_g$. We have set $\mu_{\rm mess}=10^{12}$ GeV, $n_Q=n_U = n_D = n_L = n_E = $1/2, and N=3. In the order of increasing dash length the lines correspond to $M_1$ (solid), $M_2$, $M_3$. In (a) Higgs mass is in the allowed range on the left hand side of the dashed horizontal line.}
\label{mgaugino}

\end{figure}

In Figs.~\ref{mscalar} and \ref{mgaugino} we have plotted the first and the third generation scalar mass parameters and the gaugino mass parameters as a function of $\alpha_g$ for $N=3$. We note that the difference of the mass scale of $m_{\tilde{e}_1}$ and $m_{\tilde{L}_1}$ to the rest of the scalars is a good indicator for the magnitude of $\alpha_g$, with large difference implying a larger absolute value for $\alpha_g$. For $\alpha_m=1$ and $M_0=2$ TeV only $\alpha_g$ close to -1 is allowed by the Higgs mass limit (represented here by the vertical dashed line).
In such a case the squark masses turn out to be several TeVs while slepton masses can be of the order of one TeV.
Similarly, the ratios of the gaugino mass parameters $M_3$ to $M_1$ and $M_3$ to $M_2$ correspond the value of $\alpha_g$ with a large ratios implying a $\alpha_g$ close to -1.

\section{Renormalization group invariants in deflected mirage mediation}
\label{sec:rginvariants}
Renormalization group invariants are linear combinations of the soft mass parametersq that remain 
constant under one-loop renormalization group running. Complete renormalization group invariants for the MSSM together with corresponding sum rules 
have been derived in \cite{Carena:1996km}. The RGIs are derived under several assumptions, including the vanishing of first and 
second generation Yukawas\cite{Carena:1996km}. We have derived the invariants 
for general $b_i$ in order to determine their values above the messenger scale(Table \ref{inv2}). 
Since in deflected mirage mediation new particles are introduced at a certain renormalization scale and integrated out below, one should verify whether the RGIs are affected by the modified spectrum.
By considering the renormalization group beta functions (with $\beta(p)\equiv16\pi^2\frac{dp}{dt}$ and $t=\log(\mu/\mu_0)$) for the gauginos and for the coupling constants $(a=1,2,3)$,
\begin{eqnarray}
\beta(g_a) &=& b_ag_a^3 , \label{gbeta}\\
\beta(M_a) &=& 2b_ag_a^2M_a,  \label{gauginobeta}
\end{eqnarray}
and by noting that $\beta(\frac{M_r}{g_r^2})=0$, we can define a quantity that is constant under renormalization group evolution,

\begin{equation}
I_{B_r}\equiv\frac{M_r}{g_r^2}.
\label{IBR}
\end{equation}
Similarly from the full set of MSSM renormalization group equations one can define 12 other invariants that we have enumerated in Tables \ref{inv2} and \ref{coef}. If the messenger fields would not enter the theory  at an energy scale different of the GUT-scale, the invariants would remain unchanged at all scales. Since the appearance of the messengers modifies the beta functions and contributes to gaugino and scalar masses at the messenger scale $\mu_{\rm mess}$, the invariant can have a different value above and below the scale.

Just above the messenger scale the value is equal to the GUT-scale value
\begin{equation}
I_{B_r}(\mu_{\rm mess}^+)=\frac{M_r(\mu_{\rm mess}^+)}{g_r^2(\mu_{\rm mess})}=I_{B_r}(\mu_{\rm GUT}).\label{imgutmes}
 \end{equation}
We define $\mu_{\rm mess}^+$ as evaluation at $\mu_{\rm mess}$ with modified coefficients $b_a'$ and without the threshold corrections added to gaugino and scalar masses, and  $\mu_{\rm mess}^-$ as evaluation at  $\mu_{\rm mess}$ with the usual MSSM coefficients $b_a$ and with threshold corrections added.
Below the messenger scale the gaugino masses receive the threshold correction (\ref{deltaM}). Consequently just below the messenger scale
\begin{equation}
I_{B_r}(\mu_{\rm mess}^-)=\frac{M_r(\mu_{\rm mess}^-)}{g_r^2(\mu_{\rm mess})}=\frac{M_r(\mu_{\rm mess}^+)}{g_r^2(\mu_{\rm mess})}+\frac{\Delta M_r}{g_r^2(\mu_{\rm mess})}=I_{B_r}(\mu_{\rm GUT})+\Delta I_{B_r},
\label{IBRmumess}
 \end{equation}
where $\Delta M_r$ is as in (\ref{deltaM}) and we have defined
\begin{equation}
\Delta I_{B_r}\equiv\Delta M_r/g_r^2(\mu_{\rm mess})=-NM_0/(16 \pi^2)(1+\alpha_g)\alpha_m \ln \frac{M_P}{m_{3/2}}.
\label{deltaIBR}
 \end{equation}
We evolve the couplings down from the GUT scale and obtain
\begin{eqnarray}
g_1(\mu_{\rm mess})=\frac{2\sqrt{10}\pi}{\sqrt{40\pi^2/g^2+(33+5N)t_{\text{mess}}}} ,\nonumber \\
g_2(\mu_{\rm mess})=\frac{2\sqrt{2}\pi}{\sqrt{8\pi^2/g^2+(1+N)t_{\text{mess}}}} ,\label{gmess}\\
g_3(\mu_{\rm mess})=\frac{2\sqrt{2}\pi}{\sqrt{8\pi^2/g^2+(N-3)t_{\text{mess}}}},\nonumber
\end{eqnarray}
where $t_{\text{mess}}=\ln{\mu_{\rm GUT}/\mu_{\rm mess}}$.
Thus while the invariants remain invariant from messenger scale to GUT scale as well as from the eletroweak scale to messenger scale, there is a discontinuity at the messenger scale, which must be taken into account, unless the threshold contributions are cancelled out. 

The invariants designated $D_I$ are linear combinations of the squared scalar masses with the GUT-scale value of the from
\begin{equation}
D_{I}=\gamma_I m_{3/2}^2+ \delta_I M_0^2.
\end{equation}
The invariants are constructed in such a way that threshold corrections (\ref{deltam}) cancel at the messenger scale, thus $D_I(\mu_{\text{TeV}})=D_I(\mu_{\text{GUT}})$, and they are "true" 
invariants in models that include gauge mediated supersymmetry breaking such as the deflected mirage mediation. We present $\dx$ as an example with the GUT-scale value
\begin{equation}
D_{\chi_1}\equiv3[3m_{\tilde{d}_1}^2-2(m_{\tilde{Q}_1}^2-m_{\tilde{L}_1}^2)-m_{\tilde{u}_1}^2]-m_{\tilde{e}_1}^2=M_0^2 (5+3n_U -9n_D-6n_L+n_E +6n_Q).
\label{dkhieq1}
 \end{equation}
We use the scalar mass based invariants $D_I$ to derive the high energy parameters of the deflected mirage mediation in terms of the scalar masses. 
The three invariants $I_{M_a}$ are linear combinations of the squares of both scalar and gaugino masses, and are also explicitly dependent on $b_a$, e.g. 
\begin{equation}
I_{M_1}=M_1^2-\frac{5b_1}{8}(m_{\tilde{d}_1}^2-m_{\tilde{u}_1}^2-m_{\tilde{e}_1}^2),
\end{equation}
where $b_1$ is replaced by $b_1'=b_1+N$ above the messenger scale. Thus the shift at the messenger scale has a  complex form,
\begin{align}
\Delta I_{M_1}\equiv& I_{M_1}(\mu_{\rm mess}^+)-I_{M_1}(\mu_{\rm mess}^-)\nonumber\\=&(M_1+\Delta M_1)^2-\frac{5b_1}{8}\left((m_{\tilde{d}_1}^2+\Delta m_{\tilde{d}_1}^2)-(m_{\tilde{u}_1}^2+\Delta m_{\tilde{u}_1}^2)-(m_{\tilde{e}_1}^2+\Delta m_{\tilde{e}_1}^2)\right)-M_1^2+\frac{5b_1'}{8}(m_{\tilde{d}_1}^2-m_{\tilde{u}_1}^2-m_{\tilde{e}_1}^2)\nonumber\\=&-2M_1\Delta M_1-\Delta M_1^2+\frac{5N}{8}(m_{\tilde{d}_1}^2-m_{\tilde{u}_1}^2-m_{\tilde{e}_1}^2)-\frac{5b_1}{8}(\Delta m_{\tilde{d}_1}^2-\Delta m_{\tilde{u}_1}^2-\Delta m_{\tilde{e}_1}^2), \label{deltaim1}
\end{align}
where the gaugino mass and the scalar squareds are evaluated at $\mu_{\rm mess}^+$ and $\Delta M_r$ and $\Delta m_i$ are defined in (\ref{deltaM}) and (\ref{deltam}). Since $\Delta I_{M_i}$ depends on masses at the messenger scale, accessing GUT-scale values from the TeV scale measurements is not as straightforward as with $D_I$. As with other invariants, $\Delta I_{M_r}$ is generated by the messengers and vanishes when the messengers are removed with $N=0$.

\begin{figure}
 
 \includegraphics[height=10cm]{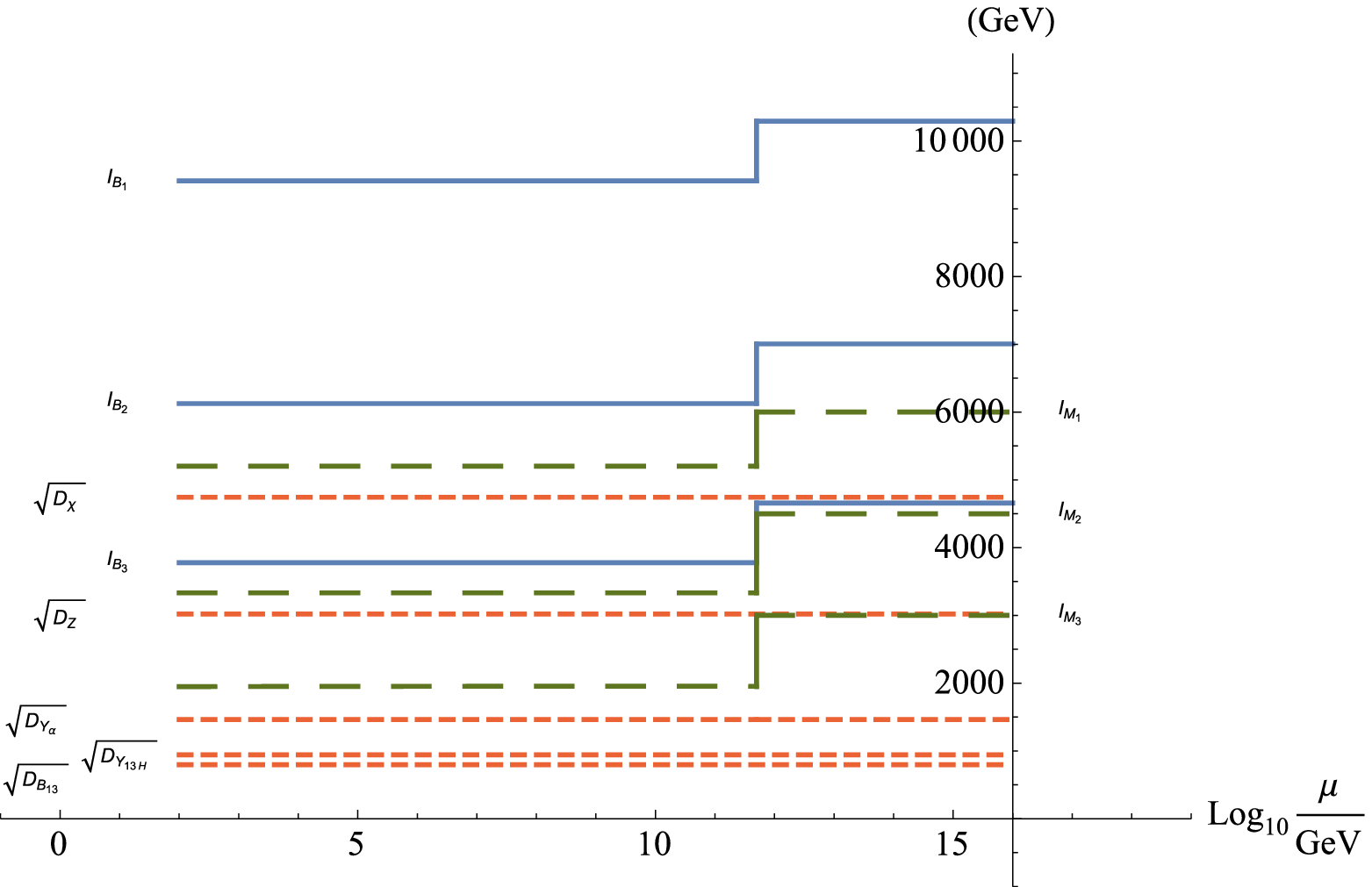}

\caption{The renormalization group invariants $\ib{a}$ (blue, solid) and the square roots of the absolute value of the invariants $D_I$ (red, small dash) and $\im{a}$ (green, large dash). Here $n_u=1/2$, $n_h=$, $M_0=3$ TeV $\alpha_m=1$, $\alpha_g=-0.5$, $\mumes=10^{12}$ GeV, and $N=3$. $D_{L_{13}}=0$ is not shown.}
\label{InvFig}

\end{figure}

We have listed  the correction at the messenger scale and the value at the GUT-scale for each invariant in Table \ref{inv2}, except for
the $D_I$ invariants which are listed in Table \ref{coef}. Fig. \ref{InvFig}. shows the values of the invariants $\ib{a}$ and the square roots of $\im{a}$, and $D_I$ above and below the messenger scale at the point $M_0=3$ TeV, $N=3$, $\alpha_m=1$, $\alpha_g=-0.5$, $\tmes=-10$, $n_u=1/2$, and $n_h=1$.

\begin{table}
\centering
\small
\begin{tabular}{|c|l|l|l|}

\hline
&&&\\
Invariant & Definition & Correction at the messenger scale & Value at the GUT scale\\
 &&&  \\
\hline
\hline
&&&\\
$I_{B_r}$&$M_r/g_r^2$&$\Delta M_r/g_r^2$&$M_0/g^2 + \frac{b_a'}{16\pi^2}m_{3/2}$\\
&&&\\

\hline
&&&\\
$I_{M_{1}}$&$M_1^2-\frac{5b_1}{8}(m_{\tilde{d}_1}^2-m_{\tilde{u}_1}^2-m_{\tilde{e}_1}^2)$&$\begin{array}{l}-2M_1\Delta M_1 - \Delta M_1^2+\\\frac{5N}{8}(m_{\tilde{d}_1}^2-m_{\tilde{u}_1}^2-m_{\tilde{e}_1}^2) \\- \frac{5b_1}{80}\Delta m ^2g_{1}^4\end{array}$&$M_0^2(1+\nepsilon{1}b_1')$\\
&&&\\
\hline
&&&\\
$I_{M_{2}}$&$M_2^2+\frac{b_2}{24}\left(9(m_{\tilde{d}_1}^2-m_{\tilde{u}_1}^2)+16m_{\tilde{L}_1}^2-m_{\tilde{e}_1}^2\right)$&$\begin{array}{l}-2M_2\Delta M_2 - \Delta M_2^2-\\ \frac{N}{24}\left(9(m_{\tilde{d}_1}^2-m_{\tilde{u}_1}^2)+16m_{\tilde{L}_1}^2-m_{\tilde{e}_1}^2\right)\\+\frac{3b_2}{48}\Delta m^2 g_{2}^4 \end{array}$&$M_0^2(1+\nepsilon{2}b_2')$\\
&&&\\
\hline
&&&\\
$I_{M_{3}}$&$M_3^2 + \frac{b_3}{16}(5m_{\tilde{d}_1}^2+m_{\tilde{u}_1}^2-m_{\tilde{e}_1}^2)$&$\begin{array}{l}2M_3\Delta M_3+ \Delta M_3^2-\\ \frac{N}{16}(5m_{\tilde{d}_1}^2+m_{\tilde{u}_1}^2-m_{\tilde{e}_1}^2)\\+\frac{b_3}{16}\Delta m^2g_{3}^4 \end{array}$&$M_0^2(1+\nepsilon{3}b_3')$\\
&&&\\
\hline
&&&\\
$I_{g_2}$&$1/g_1^2- (b_1/b_2) g_2^{-2}$&$28N/(5g_2^2(1+N))$&$1/g^2\left(1- b_1'/b_2'\right)$\\ 
&&&\\
\hline
&&&\\
$I_{g_3}$&$1/g_1^2-(b_1/b_3) g_3^{-2}$&$-16N/(5g_3^2(3-N))$&$1/g^2\left(1-b_1'/b_3'\right)$\\
&&&\\
\hline
\end{tabular}
\caption{The renormalization group invariants $I_A$ involving scalar masses, gaugino masses and coupling constants. The second column defines the invariant in terms of soft masses and couplings without messenger fields present. The third column describes the difference of the value of the invariant above and below messenger scale; the masses and the couplings are to be evaluated at the messenger scale. The fourth column describes the value of the invariant at the GUT scale; the couplings are to be evaluated at the GUT scale. The quantity $\Delta m^2$ is defined as $N/(16\pi^4) \left(M_0\alpha_m  (1+\alpha_g)  \ln \frac{M_P}{m_{3/2}}\right)^2$.The combinations of modular weights $\nepsilon{a}$ are defined in \rf{nepsilon1}-\rf{nepsilon3}.}
\label{inv2}

\end{table}

\subsection{Solving parameters using the invariants.}

We will now attempt to utilize the RGIs to solve high scale parameters $\mgr$, $M_0$, the messenger scale $\mu_{\text{mess}}$, and the number of messenger pairs $N$ in terms of low energy masses and mass parameters.

Following (\ref{IBRmumess}) we can write three equations involving the invariants $I_{B_r}$ by setting the low energy scale value of $I_{B_r}$ equal to the value at the GUT-scale corrected by the difference at the messenger scale,  
\begin{equation}
I_{B_a}(\mu_{\rm GUT})=I_{B_a}(\mu_{\rm TeV})-\Delta I_{B_a}, \label{IBReq}
\end{equation}
where $I_{B_r}$ is defined in (\ref{IBR}) and $\Delta I_{B_r}$ in (\ref{deltaIBR}). We note that $\Delta \ib{r}$ vanishes if $\alpha_g=-1$. Thus the equivalence of the TeV-scale value of $\ib{a}$ to its GUT-scale value cannot be taken as proof of the absence of gauge messengers. 

The equations (\ref{IBReq}) provide three independent solutions for $\mgr$, and $\alpha_g$, which we distinguish from each other by designating with the subindex $(a)$,
\begin{align}
&{\mgr}_{(a)}=16\pi^2\frac{\sum_{b,c=1}^3\epsilon_{abc}\ib{c}}{\sum_{d,e=1}^3\epsilon_{ade}b_e}, \qquad  (a=1,2,3), \label{m32sol}\\
&{\alpha_g}_{(a)}=\frac{b_a g^2 \mgr-16 \pi ^2 g^2 \ib{a}+16 \pi ^2 M_0}{g^2 \mgr N},  \qquad  (a=1,2,3). \label{alphagsol}\\
\end{align}
It is easy to verify by evolving $g_i$ that the deflected mirage mediation coupling constant at the GUT scale, $g$, is related to $g_{\text{GUT}}$, which is the coupling constant at the GUT-scale with $N=0$, by
\begin{equation}
\frac{1}{g}=\sqrt{\frac{1}{g_{\text{GUT}}^2}-\frac{N t_{\text{mess}}}{8 \pi ^2}}.\label{gdef}
\end{equation}
A different set of parameters to eliminate could, of course, be chosen, but this choice proves to be most convenient for solving all parameters, as $M_0$ is readily solved from the scalar mass involving invariants $D_I$ which do not allow the determination of $\alpha_g$.
\begin{table}
\centering
\begin{tabular}{|c|c|c|c|}
\hline
&&&\\
I&Definition &$\gamma_I$&$\delta_I$\\
&&& \\
\hline
\hline
&&&\\
$D_Z$&$3(m_{\tilde{d}_3}^2-m_{\tilde{d}_1}^2)+2(m_{\tilde{L}_3}^2-m_{H_d}^2)$&$Y_{Za}+Y_{Zb}g^2$&$2n_{\alpha}$\\
&&&\\
\hline
&&&\\
$D_{\chi_1}$&$3(3m_{\tilde{d}_1}^2-2(m_{\tilde{Q}_1}^2-m_{\tilde{L}_1}^2)-m_{\tilde{u}_1}^2)-m_{\tilde{e}_1}^2$&0&$n_{\beta}$ \\
&&&\\
\hline
&&&\\
$D_{L_{13}}$&$2(m_{\tilde{L}_1}^2-m_{\tilde{L}_3}^2)-m_{\tilde{e}_1}^2+m_{\tilde{e}_3}^2$&$0$&$0$\\
&&&\\
\hline
&&&\\
$D_{B_{13}}$&$2(m_{\tilde{Q}_1}^2-m_{\tilde{Q}_3}^2)-m_{\tilde{u}_1}^2+m_{\tilde{u}_3}^2-m_{\tilde{d}_1}^2+m_{\tilde{d}_3}^2$&$Y_{B_{13}a}+Y_{B_{13}b}g^2$&0\\
&&&\\
\hline
&&&\\ 
$D_{Y_{13H}}$&$\begin{array}{c} m^2_{\tilde{Q}_1}-2m^2_{\tilde{u}_1}+m^2_{\tilde{d}_1}-m^2_{\tilde{L}_1}+m^2_{\tilde{e}_1}\\-\frac{10}{13}\left(m^2_{\tilde{Q}_3}-2m^2_{\tilde{u}_3}+m^2_{\tilde{d}_3}-m^2_{\tilde{L}_3}+m^2_{\tilde{e}_3}+m^2_{H_u}-m^2_{H_d}\right)\end{array}$&$\frac{10}{13}\left(-Y_{\alpha 1}-Y_{\alpha 2}g^2\right)$&$-\frac{1}{13}n_{\gamma}$\\
&&&\\
\hline
&&&\\
$D_Y\alpha$&$\begin{array}{l}\left(m^2_{H_u}-m^2_{H_d}+\sum_{gen}(m^2_{\tilde{Q}}-2m^2_{\tilde{u}}+m^2_{\tilde{d}}-m^2_{\tilde{L}}+m^2_{\tilde{e}})\right)/g^2\end{array}$\ &$\frac{1}{g^2}\left(Y_{\alpha 1}+Y_{\alpha 2}\right)$&$-\frac{1}{g^2}n_{\delta}$\\
&&&\\
&&&\\
\hline
\end{tabular}
\caption{The invariants $D_{I}$ and their GUT-scale values parametrised as $D_I(\mu_{\text{GUT}})=\gamma_I m_{3/2}^2+ \delta_I M_0^2$ . See (\ref{nY1})-(\ref{nY9}).} \label{coef}
\end{table}

The invariants $\dx$, $D_{B_{13}}$, $D_{Y_{13H}}$, $D_Z$, and $D_Y\alpha$ are composed of linear combinations of soft scalar mass parameters and have identical values above and below the messenger scale as the linear combinations of the mass parameters are chosen so that the threshold corrections (\ref{deltam}) cancel out.
The invariants have the schematic form
\begin{equation}
D_{I}=\gamma_I m_{3/2}^2+ \delta_I M_0^2,
\end{equation}
at the GUT-scale, where the coefficients $\gamma_I$, $\delta_I$, $\epsilon_I$, are determined by the Yukawa couplings $y_{\tau}$, $y_{t}$ and $y_{b}$ and $g$ at the GUT scale We have listed the values in Table \ref{coef}, where
\begin{align}
n_{\alpha}\equiv&n_{H_d}-n_L,\\
n_{\beta}\equiv&5+3n_U -9n_D-6n_L+n_E +6n_Q,\\ 
n_{\gamma}\equiv&3 n_D + 3 n_E + 10 n_{H_d} - 10 n_{H_u} - 3 n_L + 3 n_Q - 6 n_U,\label{nY1}\\
n_{\delta}\equiv& 3 n_D + 3 n_E - n_{H_d} + n_{H_u} - 3 n_L + 3 n_Q-6 n_U,\label{nY2}\\
n_{\epsilon}\equiv& n_{Hu}+n_Q+n_U-3,\label{nY3}\\
Y_{B_{13}a}\equiv&\frac{1}{10240 \pi ^4}[1440 y_b^6+240 y_b^4 y_{\tau}^2+240 y_b^4 y_t^2-1440 y_b^4-240 y_b^2 y_{\tau}^2+240 y_b^2 y_t^4-480 y_b^2 y_t^2\nonumber\\&+1440 y_t^6-1440 y_t^4],\label{nY4}\\
Y_{B_{13}b}\equiv& \frac{1}{30720 \pi ^4}[-2105 y_b^4+2105 y_b^2-2208 y_t^4+2208 y_t^2],\label{nY5}\\
Y_{Za}\equiv&\frac{1}{10240 \pi ^4}[1440 y_b^6+240 y_b^4 y_{\tau}^2+240 y_b^4 y_t^2+1440 y_b^4-720 y_b^2 y_{\tau}^2-240 y_b^2 y_t^2-640 y_{\tau}^4],\label{nY6}\\
Y_{Zb}\equiv&\frac{1}{10240 \pi ^4}[-2105 y_b^4+2105 y_b^2+768 y_{\tau}^2],\label{nY7}\\
Y_{\alpha 1}\equiv& \frac{1}{30720 \pi ^4}[1440 y_b^6+240 y_b^4 y_{\tau}^2+240 y_b^4 y_t^2-1440 y_b^4-960 y_b^2 y_{\tau}^2-480 y_b^2 y_t^4\nonumber\\&+240 y_b^2 y_t^2-960 y_{\tau}^4-2880 y_t^6+2880 y_t^4],\label{nY8}\\
Y_{\alpha 2}\equiv&\frac{1}{30720 \pi ^4}[-2105 y_b^4+2105 y_b^2+1152 y_{\tau}^2+4416 y_t^4-4416 y_t^2].\label{nY9}
\end{align}

The parameter $M_0$ is easily solved using the GUT-scale value of $\dx$,
\begin{align}
 &D_{\chi_1}=3\left(3m_{\tilde{d}_1}^2-2(m_{\tilde{Q}_1}^2-m_{\tilde{L}_1}^2)-m_{\tilde{u}_1}^2\right)-m_{\tilde{e}_1}^2 =n_{\beta}M_0,
\label{DZdx}
\end{align}
to obtain
\begin{align}
&M_0= \sqrt{\frac{\dx}{n_{\beta}}}, \text{for }n_{\beta}\neq0,
\label{M0sol}
\end{align}
For solving $g$ we use the GUT-scale values of the two invariants $D_{Y_{13H}}$, $D_Y\alpha$, defined as
\begin{align}
D_{Y_{13H}}=&m^2_{\tilde{Q}_1}-2m^2_{\tilde{u}_1}+m^2_{\tilde{d}_1}-m^2_{\tilde{L}_1}+m^2_{\tilde{e}_1}-\frac{10}{13}\left(m^2_{\tilde{Q}_3}-2m^2_{\tilde{u}_3}+m^2_{\tilde{d}_3}-m^2_{\tilde{L}_3}+m^2_{\tilde{e}_3}+m^2_{H_u}-m^2_{H_d}\right)\\\nonumber=&\frac{10}{13}\left(-Y_{\alpha 1}-Y_{\alpha 2}g^2\right)\mgr^2-\frac{1}{13}n_{\gamma}M_0^2,\\
D_Y\alpha=&\left(m^2_{H_u}-m^2_{H_d}+\sum_{gen}(m^2_{\tilde{Q}}-2m^2_{\tilde{u}}+m^2_{\tilde{d}}-m^2_{\tilde{L}}+m^2_{\tilde{e}})\right)/g^2=\frac{1}{g^2}\left(Y_{\alpha 1}+g^2Y_{\alpha 2}\right)\mgr^2-\frac{1}{g^2}n_{\delta}M_0^2,
\end{align}
Yukawa-dependent terms $Y_{\alpha 1}$ and $Y_{\alpha 1}$ can be eliminated by forming the linear combination

\begin{align}
\dy+\frac{10g^2}{13}\dya=-\frac{10}{13}M_0^2 ( n_{\delta}+n_{\gamma}). \label{dydya}
\end{align}
We obtain the solution 
\begin{align}
g^2=& -\frac{\mathcal{Y}(n_{\delta\gamma})}{10 \dya},\label{gsol}\\
\end{align}
where we have defined 
\begin{align}
&n_{\delta\gamma}=n_{\delta}+n_{\gamma}=2n_{\gamma}-11n_{H_d}+11n_{H_d},\\
&\mathcal{Y}(n_A)=13 D_{Y_{13H}}+10M_0^2n_A.\\
\end{align}
Using the solution \rf{M0sol} for $M_0$, we write
\begin{align}
&g^2=\frac{-10 \dx (n_{\delta}+n_{\gamma})-13 \dy n_{\beta}}{10\dya \sqrt{n_{\beta}}}\label{gsol1},\\
&\mathcal{Y}(n_A)=13\dy+\frac{10n_A}{\nbeta}\dx .
\end{align}
Since from (\ref{gdef}) we have
\begin{align}
\tmes= \frac{8\pi ^2 \left( g^2- \gGUT^2\right)}{g^2 \gGUT^2 N}, \label{tmessol}
\end{align}
the relation of $\tmes$ and $N$ is fixed once $g^2$ is measured. In addition to (\ref{gsol}), equations 
\begin{equation}
\begin{cases}
\dy&=\frac{10}{13}\left(-Y_{\alpha 1}-Y_{\alpha 2}g^2\right)\mgr^2-\frac{1}{13}n_{\gamma}M_0^2,\\
\dya&=\frac{1}{g^2}\left(Y_{\alpha 1}+Y_{\alpha 2}\right)\mgr^2-\frac{1}{g^2}n_{\delta}M_0^2,
\end{cases}\label{dydya}
\end{equation}
produce a solution to $\mgr$,
\begin{align}
\mgr=\frac{32 \sqrt{30} \pi ^2 \sqrt{\dya \mathcal{Y}(n_{\gamma})}}{\sqrt{Y_{\alpha 2} \mathcal{Y}(n_{\delta\gamma})-10 \dya Y_{\alpha 1}}}.\label{m32sol1}
\end{align}
By solving simultaneously the equations 
\begin{equation}
\begin{cases}
\db&=(Y_{B_{13}a}+Y_{B_{13}b}g^2)\mgr^2\\
\dz&=2n_{\alpha} M_0^2+(Y_{Za}+Y_{Zb}g^2)\mgr^2
\end{cases},\label{dbdz}
\end{equation}
we find alternative solutions to $\M_0$ and $\mgr$ that are dependent on the Yukawa couplings 
\begin{align}
\mgr=&\frac{32 \sqrt{30} \sqrt{\db}}{\sqrt{g^2 Y_{B_{13}b}+Y_{B_{13}a}}}\rv=\frac{320 \sqrt{3} \sqrt{\db} \sqrt{\dya} \sqrt{n_{\beta}}}{\sqrt{10 \dya n_{\beta} Y_{B_{13}a}-Y_{B_{13}b} \mathcal{Y}(n_{\delta\gamma})}},\label{m32sol2}\\
M_0=&\frac{\sqrt{3 \db \left(g^2 Y_{Zb}+Y_{Za}\right)-\pi ^4 \dz \left(g^2 Y_{B_{13}b}+Y_{B_{13}a}\right)}}{\sqrt{2} \pi ^2 \sqrt{-n_{\alpha} \left(g^2 Y_{B_{13}b}+Y_{B_{13}a}\right)}}=\rv \sqrt{\frac{\left(\pi ^4 \dz Y_{B_{13}b}-3 \db Y_{Zb}\right)\mathcal{Y}(n_{\delta\gamma})-10 \dya n_{\beta} \left(\pi ^4 \dz Y_{B_{13}a}-3 \db Y_{Za}\right)}{2 \pi^4 n_{\alpha}\left(Y_{B_{13}b} (\mathcal{Y}(n_{\delta\gamma})-10 \dya n_{\beta} Y_{B_{13}a}\right)}}.\label{M0sol1}
\end{align}

\subsubsection{Yukawas in terms of invariants}
Instead of the mass parameters it is possible to use $D_I$ to determine the values of the Yukawas at the GUT scale, although the necessary parameters for running the Yukawas up from the low energy scale are solvable independently of the Yukawas. We assume a small $y_b$ and $y_{\tau}$ compared to $y_t$ and small enough $y_t$ to neglect terms with $y_t^6$ and higher order, and $y_b^4$, $y_{\tau}^4$ and higher order. The equations (\ref{nY4})-(\ref{nY9}) are then reduced to
\begin{align}
Y_{B_{13}a}=&-1440 y_t^4,\\
Y_{B_{13}b}=&-2105 y_b^2-2208 y_t^4+2208 y_t^2,\\
Y_{Za}=&0,\\
Y_{Zb}=&2105 y_b^2,\\
Y_{\alpha 1}=& 2880 y_t^4,\\
Y_{\alpha 2}=&2105 y_b^2+1152 y_{\tau}^2+4416 y_t^4-4416 y_t^2.
\end{align}
From (\ref{dydya}), and (\ref{dbdz}) we can then solve 
\begin{align}
y_{\tau}=&\frac{4  \sqrt{5} \sqrt{-\dya} \sqrt{60 \db-\pi ^4 \left(13\Ycal{\ngamma}-10 \dz+20 M_0^2 \nalpha\right)}}{3 \sqrt{\mgr^2 \Ycal{\ndg}}},\label{ytausol}\\
y_{b}=&\frac{64 \sqrt{\frac{5}{421}} \pi ^2 \sqrt{-\dya \left(\dz-2 M_0^2 \nalpha\right)}}{\sqrt{\mgr^2 \Ycal{\ndg}}},\label{ybsol}\\
y_t=&\frac{1}{6\mgr^2 \left(23\Ycal{\ndg}-150 \dya\right)}\bigg\{4761 \mgr^4 \Ycal{\ndg}^2+69\Ycal{\ndg}\rv+\sqrt{38400 \dya \mgr^2 \left(3 \db+\pi ^4 \left(\dz-2 M_0^2 \nalpha\right)\right) \left(23\Ycal{\ndg}-150 \dya\right)}\bigg\}.\label{ytsol}\\
\end{align}
Note that while the solution $\dx/\nbeta$ can be substituted for $M_0^2$, and $16 \pi ^2 (I_{B_1}-I_{B_3})/(b_1-b_3)$ for $\mgr$ here, the solution (\ref{M0sol1}) for $M_0$ and the  solutions (\ref{m32sol1}) and (\ref{m32sol2}) for $\mgr$ derived above, are not independent from (\ref{ytausol})-(\ref{ytsol}) and thus cannot be used.

\subsubsection{Solving N} 
For solving $N$ we use the remaining three invariants composed of the mass parameters,
\begin{align}
&I_{M_1}=M_1^2-\frac{5b_1'}{8}(m_{\tilde{d}_1}^2-m_{\tilde{u}_1}^2-m_{\tilde{e}_1}^2),\label{im1def2}\\
&\im{2}=M_2^2+\frac{b_2'}{24}\left(9(m_{\tilde{d}_1}^2-m_{\tilde{u}_1}^2)+16m_{\tilde{L}_1}^2-m_{\tilde{e}_1}^2\right),\\
&\im{3}=M_3^2 + \frac{b_3'}{16}(5m_{\tilde{d}_1}^2+m_{\tilde{u}_1}^2-m_{\tilde{e}_1}^2).
\end{align}
We remind the reader that parameter $b_a'=b_a+N$ is to be replaced with $b_a$ below the messenger scale.
In order to connect TeV scale measurements to the GUT scale parameter values, we examine $\im{1}$ above the messenger scale:
\begin{align}
I_{M_1}(\muplus)=&M_1^2(\muplus)-\frac{5b_1'}{8}(m_{\tilde{d}_1}^2(\muplus)-m_{\tilde{u}_1}^2(\muplus)-m_{\tilde{e}_1}^2(\muplus))\nonumber\\ =&I_{M_1}(\mugut).
\end{align}
By using the relations
\begin{align}
&\ib{1}(\mugut)=\ib{1}(\muplus)=M_1(\muplus)/\gmes{1}{4},\label{ib1eq1}\\
 &m_i^2(\muplus)=m_i^2(\muminus)-\Delta m_i^2, \nonumber
 \end{align}
we eliminate the mass parameters measured above messenger scale and obtain
\begin{align}
I_{M_1}(\mugut)=&\iba^2(\mugut)g_1^4(\mumes)-\frac{5b_1'}{8}(m_{\tilde{d}_1}^2(\muminus)-m_{\tilde{u}_1}^2(\muminus)-m_{\tilde{e}_1}^2(\muminus))\nonumber\\ +&\frac{5b_1'}{8}(\Delta m_{\tilde{d}_1}^2+\Delta m_{\tilde{u}_1}^2-\Delta m_{\tilde{e}_1}^2).
\end{align}
Directly from (\ref{im1def2}) and (\ref{ib1eq1}) we see that  below the messenger scale $5b_1/8(m_{\tilde{d}_1}^2-m_{\tilde{u}_1}^2-m_{\tilde{e}_1}^2)=\im{1}-M_1^2=\im{1}-\ib{1}\gmes{1}{4}$. Thus
\begin{align}
I_{M_1}(\mugut)=&\iba^2(\mugut)g_1^4(\mumes)+\frac{b_1'}{b_1}\left(\ima-\iba^2g_1^4(\mumes)\right)\nonumber\\ +&\frac{5b_1'}{8}(\Delta m_{\tilde{d}_1}^2+\Delta m_{\tilde{u}_1}^2-\Delta m_{\tilde{e}_1}^2).\label{im1gut}
\end{align}
We have dropped the argument $\muminus$ so that the low scale value of the invariants are meant unless otherwise specified.
Analoguously
\begin{align}
I_{M_2}(\mugut)=&\ibb^2(\mugut)g_2^4(\mumes)+\frac{b_1'}{b_1}\left(\ima-\ibb^2g_2^4(\mumes)\right)\nonumber\\ -&\frac{b_2'}{24}\left(9(\Delta m_{\tilde{d}_1}^2-\Delta m_{\tilde{u}_1}^2)+16\Delta m_{\tilde{L}_1}^2-\Delta m_{\tilde{e}_1}^2\right),\label{im2gut}\\
I_{M_3}(\mugut)=&\ibc^2(\mugut)g_3^4(\mumes)+\frac{b_1'}{b_1}\left(\imc-\ibc^2g_3^4(\mumes)\right)\nonumber\\ -&\frac{b_3'}{16}(5\Delta m_{\tilde{d}_1}^2+\Delta m_{\tilde{u}_1}^2-\Delta m_{\tilde{e}_1}^2)\label{im3gut}.
\end{align}
After substituting the GUT scale values $\iba(\mugut)$ and $\ima(\mugut)$, the corrections to the scalar masses (\ref{deltam}) with $\ag$ from (\ref{alphagsol}), $g_a(\mumes)$ from (\ref{gmess}), and ${\ag}_{(a)}$ , the equations (\ref{im1gut})-(\ref{im3gut}) can be collectively written as 

\begin{align}
&256 \pi ^4 \bigg\{-b_a^2(g^2-\gGUT^2)^2 \left(\mathcal{I}_{M_a}-M_0^2\right)+b_a g^2 N \left(g^2 \left(M_0^2-2 \mathcal{I}_{M_a}\right) \gGUT^2 \mathcal{I}_{M_a}\right)-g^4 \mathcal{I}_{M_a} N^2\bigg\}\rv+g^4 \mathcal{I}_{B_a}^2 N (b_a+N)+32 \pi ^2 b_a g^2 \gGUT^2 \mathcal{I}_{B_a} M_0 N=0,\qquad (a=1,2,3),\label{iba}\end{align}
where we have defined
\begin{align}
&\mathcal{I}_{B_a}\equiv(b_a\mgr-16 \pi ^2 I_{B_a})\gGUT^2,\label{calIB}\\
&\mathcal{I}_{M_a}\equiv I_{M_a}-b_a M_0^2n_{\epsilon_a},\label{calIM}
\end{align}
where
\begin{align}
&n_{\epsilon_1}\equiv\frac{5}{8}\left(1+n_D-n_e-n_U\right),\label{nepsilon1}\\
&n_{\epsilon_2}\equiv\frac{1}{24}\left(15-9n_D+n_e+9n_U-16n_L\right),\label{nepsilon2}\\
&n_{\epsilon_3}\equiv\frac{1}{16}\left(5-5n_D+n_e-n_U\right).\label{nepsilon3}
\end{align}

All $\mgr$-dependence in \rf{iba} is  now contained in $\ibcal{a}$, and can be eliminated with the solution ti ${\mgr}_{(a)}$ from \rf{m32sol} to obtain
\begin{align}
\ibcal{a}=16\pi^2\gGUT^2\frac{\sum_{b,c}\epsilon_{abc}\left(b_a\ib{c}-b_c\ib{a}\right)}{\sum_{d,e}\epsilon_{ade}b_e}.
\end{align}

A solution for $N$ can  be obtained from each of the equations \rf{iba},
\begin{align}
N_{(a)}&=-\frac{b_a}{2 g^2 \left(\mathcal{I}_{B_a}^2-256 \pi ^4 \mathcal{I}_{M_a}\right)}\bigg\{g^2 \left(\mathcal{I}_{B_a}^2+256 \pi ^4 \left(-2\mathcal{I}_{M_a}- M_0^2\right)\right)+32 \gGUT^2 \left(\pi ^2 \mathcal{I}_{B_a} M_0+16\pi ^4 \mathcal{I}_{M_a}\right)\rv \pm \left(\mathcal{I}_{B_a}+16 \pi ^2 M_0\right) \sqrt{g^4 \left(\mathcal{I}_{B_a}-16 \pi ^2 M_0\right)^2+64 \pi ^2 g^2 \gGUT^2 \left(\mathcal{I}_{B_a} M_0-16 \pi ^2 \mathcal{I}_{M_a}\right)+1024\pi ^4 \gGUT^4 \mathcal{I}_{M_a}}\bigg\}.\label{Nsol}
\end{align}
Solutions can also be obtained for $M_0$, $\nepsilon{a}$, $\mgr$, or $g$:
\begin{align}
{M_0}_{(a)}=&\frac{N g^2+b_a g'^2}{\nepsilon{a}(Ng^2+bg'^2)^2+bg'^4+Ng^4}\bigg\{\pm\sqrt{\left(\nepsilon{a}(Ng^2+bg'^2)^2+bg'^4+Ng^4\right)\im{a}-b_a g^4 N (\nepsilon{a} b_a'+1)\mathcal{I}_{B_a}^2}\rv -\frac{ N b_a g^2(g^2-g'^2) \mathcal{I}_{B_a}}{16 \pi ^2(N g^2+b_a g'^2)},\bigg\}\label{M0solalt}\\
\nepsilon{a}=&\frac{I_{M_a}}{b_a M_0^2}-\frac{(Ng^2\ibcal{a}-16\pi^2 b_a g'^2 M_0)^2+b_a Ng^4(\ibcal{a}+16\pi^2M_0)^2}{256 \pi ^4b_a M_0^2 \left(N g^2+b_a g'^2\right)^2},\label{nepsilonsol}\\
{\mgr}_{(a)}=&16 \pi^2 \left(\frac{\ib{a}}{b_a}-\frac{M_0}{g^2b_a'}\pm\frac{\left(Ng^2+b_a g'^2\right)\sqrt{N \left(b_a'\mathcal{I}_{M_a}-b_a M_0^2\right)}}{N b_a'b_a g^2(g^2-g'^2) }\right),\label{m32solalt}\\
g^2_{(a)}=&\pm \frac{16 \pi^2\gGUT^2b_a\left(\mathcal{I}_{M_a}-M_0^2\right)}{16 \pi ^2\left(b_a'\mathcal{I}_{M_a}-b_aM_0^2\right)+\mathcal{I}_{B_a} M_0 N\pm\left(\mathcal{I}_{B_a}+16 \pi ^2 M_0\right)\sqrt{N  \left(b_a' \mathcal{I}_{M_a}-b_a M_0^2\right)}},\label{gsolalt}
\end{align}
where
\begin{align}
b_a'=&b_a+N,\\
g'^2=&g^2-\gGUT^2.
\end{align}
The correct signs have to be determined by other means, such as by comparing solutions and ruling out negative values.
Only the solutions with a different index are independent; i.e. we cannot use the solution $N=N_{(1)}$ to solve  ${M_0}_{(1)}$ etc. Three linearly independent  solutions can be "picked", each with a different index, e.g. ${M_0}_{(1)}$, $N_{(2)}$, $g_{(3)}$, and the supplemented with the solutions obtained from $D_I$ and $\ib{a}$.

\begin{figure}
\psfrag{M}{$\ima$(GeV^2)}
\psfrag{B}{$\iba$(GeV)}
 \subfigure[N]{\psfrag{M}{$\ima$(GeV^2)}\includegraphics[height=7.0cm]{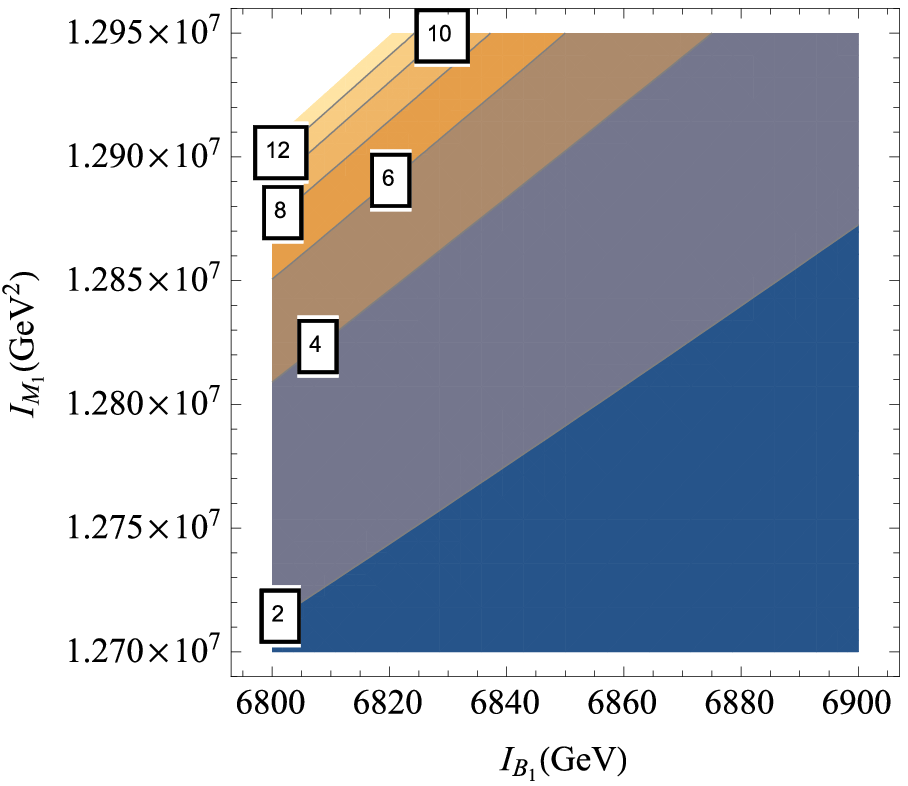}}
 \subfigure[$\alpha_g$]{ \includegraphics[height=7.0cm]{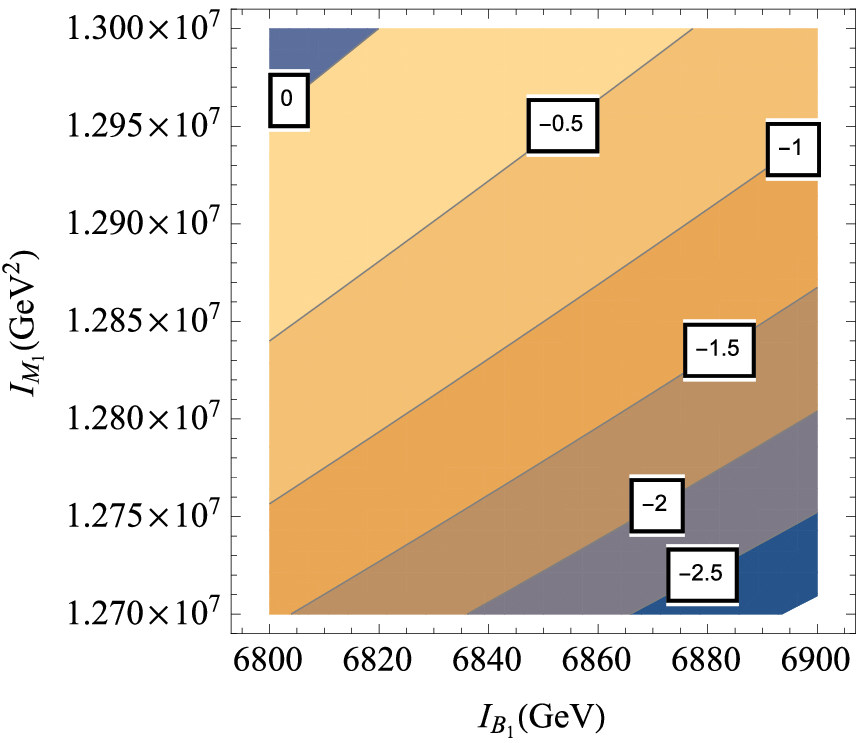}}

\caption{The number of messenger field pairs N, and $\alpha_g$ as predicted by (\ref{Nsol}) and (\ref{alphagsol}). $M_0 = 2.0$ TeV, $\alpha_m=1$, and $g=-\frac{\mathcal{Y}(n_{\delta\gamma})}{10 \dya}=0.80$, and $n_u=1/2$.}
\label{Nfig}

\end{figure}

Fig. \ref{Nfig}. illustrates the behavior of N and $\alpha_g$ as a function of $\ib{1}$ and $\im{1}$ as predicted by (\ref{Nsol}) and (\ref{alphagsol}). We assume $n_u=1/2$, and $\dya$, $\dy$, $\dx$, and two of the $\ib{a}$ to have been measured to fix $g=0.80$, $\alpha_m=1$, and $M_0=2$ TeV. Thus $\ib{1}$ and $\im{1}$ determine the values of $N$ and $\alpha_g$. Small variation in $\im{1}$ and $\ib{1}$ point to significantly different values of $N$ and $\alpha_g$; if $\im{1}=1.29\times10^7 \text{GeV}^2$, a difference of 50 GeV in $\ib{1}$ separates $N=4$ and $N=12$.  
Table \ref{inv3}. summarizes the solutions derived above. 
\subsubsection{The modular weights.}

In terms of unimodular weights $n_h=n_{H_d}=n_{H_u}$ and $n_u=n_U= n_D = n_L = n_E =n_Q$, for which
\begin{align}
&n_{\alpha}=n_h-n_u,\label{nalphayt}\\
&n_{\beta}=5(1-n_u),\label{nbetayt}\\
&n_{\gamma}=n_{\delta}=n_{\delta\gamma}=0,\label{ngammayt}\\
&\nepsilon{1}=\frac{5}{8}\left(1-n_u\right),\\
&n_{\epsilon_2}=\frac{5}{8}\left(1-n_u\right),\\
&n_{\epsilon_3}=\frac{5}{16}\left(1-n_u\right),
\end{align}
the $M_0$-term is removed from the solution of $g$, thus (\ref{gsol}) is simplified to
\begin{align}
g^2=& -\frac{13\dy}{10 \dya}.\\
\end{align}
Determining an analytical solution for $n_u$  requires solving two of the three equations (\ref{iba}) simultaneously for $N$ and $n_u$. While the solution exists, it is  too complex to 
be presented explicitly. A simpler way may be to determine $n_u$ numerically or by guessing from the equation
\begin{align}
N_{(1)}=N_{(2)}.
\end{align}
The third independent equation $N_{(3)}$ then remains for determining $N$.
$M_0$, $\mgr$, $\mumes$, and $N$ are now known, thus the Yukawas at the GUT-scale can be evaluated through conventional method of the renormalization group running. 
Then $n_h$ can be determined from 
\begin{align}
\dz=2 M_0^2 (n_h-n_u)+(Y_{Za}+Y_{Zb}g^2)\mgr^2.
\end{align}
A measurement of $\dx=0$ would present the problem of determining whether $M_0=0$ or $n_u=1$. If we assume the latter, two of the equations (\ref{iba}) could be solved simultaneously 
for $M_0$ and $N$. Then the value of  $\mgr$ from the remaining independent equation (\ref{m32solalt}) can be evaluated and checked for consistency with (\ref{m32sol}), to verify the hypothesis.

In the general case it is not possible to determine all seven combinations of the modular weights since after solving the five parameters only five independent RGIs remain. 
left. 
 Some of $\nepsilon{a}$ can be solved from (\ref{nepsilonsol}) and sum rules discussed in Section IV could be helpful in determining $\nbeta$, $\nalpha$, $\ngamma$, and $\ndelta$.

\begin{table}
\centering
\begin{tabular}{|l|c|c|c|}
\hline
&&&\\
Condition\ \ &$\ib{a}(TeV)=\ib{a}(GUT)+\Delta \ib{a}$&$D_I(TeV)=D_I(GUT)$&$\br{c}\im{a}(TeV)=\im{a}(GUT)+\Delta \im{a}; \\\rf{alphagsol}\er$\\
&&&\\
\hline
\hline
&&&\\
Solution \ \ &$\br{l}3\times\bc\mgr\ec\qquad\qquad \ \ \ (\ref{m32sol})\\3\times\bc\ag(M_0,N,\mgr)\ec(\ref{alphagsol})\er$&$\br{l} \bc M_0 \ec \qquad\qquad \ (\ref{M0sol})\\ \\ \bc \mgr(M_0)\\g(M_0)\ec \quad \br{r}(\ref{m32sol1})\\(\ref{gsol})\er\\ \\ \bc\mgr(g)\\ M_0(g)\ec \qquad\br{r}(\ref{m32sol1})\\(\ref{M0sol1})\er \\ \\\bc g(M_0) \\ y_t(M_0,\mgr)\\y_{\tau}(M_0,\mgr)\\y_b(M_0,\mgr)\ec \br{r}(\ref{gsol})\\(\ref{ytsol})\\(\ref{ytausol})\\(\ref{ybsol})\er\\ \\ \er$&$\br{l}3\times \bc  N(M_0,\mgr,g)\ec (\ref{Nsol}) \\ \\ 3\times  \bc M_0(N,\mgr,g)\ec (\ref{M0solalt})\\ \\  3\times \bc g(N,M_0, \mgr,)\ec (\ref{gsolalt})\\ \\  3\times \bc \mgr(N,M_0, g,)\ec (\ref{m32solalt})\er$\\
\hline
\end{tabular}
\caption{Summary of the solutions for the parameters derived in Chapter III.} \label{inv3}
\end{table}

\section{Sum rules in deflected mirage mediation}
\label{sec:sumrules}
Renormalization group invariants can be used to construct sum rules by applying various conditions to reduce variables, e.g. mass unification.
As a generic example we assume gaugino mass unification at some scale and write $M_1=M_2=M_3=M_{1/2}$. From (\ref{IBR}),
\begin{equation}
I_{B_a}=\frac{M_{1/2}}{g_a^2}.
\end{equation}
By combining this to the definitions of the invariants $I_{g_2}$ and $I_{g_3}$, 
\begin{eqnarray}
I_{g_2}&=&1/g_1^2- (b_1/b_2) g_2^{-2},\\
I_{g_3}&=&1/g_1^2- (b_1/b_3) g_3^{-2},
\label{IGN}
\end{eqnarray}
we can eliminate $M_{1/2}$ from the resulting group of equations to obtain the sum rule
\begin{equation}
[I_{B_1} - (b_1/b_3) I_{B_3}] I_{g_2} = [I_{B_1} - (b_1/b_2) I_{B_2}] I_{g_3}.
\label{sumgaugino}
\end{equation}
If we assume the gaugino mass unification to occur in conjunction with a scalar mass unification (with the common scalar mass squared value $m_0^2$) at the same scale, the RGIs $\im{a}$ have the values
\begin{align}
\im{1}=&\frac{33m_{0}^2}{8}+M_{1/2}^2,\\
\im{2}=&\frac{5m_{0}^2}{8}+M_{1/2}^2,\\
\im{3}=&M_{1/2}^2-\frac{15m_{0}^2}{16}.
\end{align}
This allows us to write the sum rule
\begin{align}
81 \im{2}-56 \im{3}-25\im{1}=0.
\end{align}
Similarly from the assumption of gauge coupling unification at the GUT scale one can derive 
\begin{equation}
I_{g_1}-I_{g_2}(1-b_1/b_2)/(1-b_1/b_3)=0.
\label{sumcoupling}
\end{equation}
In the specific case of deflected mirage mediation we can verify the formula easily for renormalization scales above the messenger scale by substituting $b_a'$ for $b_a$ and plugging in the GUT-scale values for the gaugino masses (\ref{gauginoUV}). 
To examine the equation below the messenger scale we restore $b_a$ and write the invariants at the scale $\mu_{\rm mess}^-$. 
Using (\ref{IGN}), (\ref{gmess}), and (\ref{IBRmumess}) it is straightforward to verify that  (\ref{sumgaugino}) is valid below  messenger scale as well.
Similarly one can derive sum rules based on scalar mass unification and scalar mass unification combined to gaugino mass unification at the same unification scale \cite{Hetzel:2012bk}.
Validity of the sum rules derived in  \cite{Hetzel:2012bk} at regions above and below the messenger scale is listed in Table~\ref{sumrule} for
deflected mirage mediation.

\begin{table}
\centering
\begin{tabular}{|c|c|c|}
\hline
&&\\
Sum rule & $\mu<\mu_{\rm mess}$ & $\mu>\mu_{\rm mess}$\\
&& \\
\hline

\hline
&&\\
$I_{g_1}-I_{g_2}(1-b_1/b_2)/(1-b_1/b_3)=0$&OK&OK\\
&&\\
\hline
&&\\
 $(I_{B_1} - (b_1/b_3) I_{B_3}) I_{g_2} = (I_{B_1} - (b_1/b_2) I_{B_2}) I_{g_3}$&OK&OK \\
&&\\
\hline
&&\\
$I_{g_2} = \left(I_{M_1}-\frac{b_1}{8}D_{\chi_1}\right)^{-1/2}I_{B_1} - \frac{b_1}{b_2}\left(I_{M_2}-\frac{b_2}{8}D_{\chi_1}\right)^{-1/2}I_{B_2}$&X&$n_i=n_u$\\
&&\\
\hline
&&\\ 
$I_{g_3} = \left(I_{M_1}-\frac{b_1}{8}D_{\chi_1}\right)^{-1/2}I_{B_1} + \frac{b_1}{b_3}\left(I_{M_3}+\frac{b_3}{16}D_{\chi_1}\right)^{-1/2}I_{B_3}$&X&$n_i=n_u$\\
&&\\
\hline
&&\\
$I_{M_1} - (2b_1+b_3)/(2b_2+b_3) I_{M_2} + 2(b_1-b_2)/(2b_2+b_3) I_{M_3}=0$&X&$n_i=1$\\
&&\\
\hline
\end{tabular}
\caption{Sum rules derived from the condition of gauge coupling unification, gaugino mass unification and scalar mass unification. Third and fourth rows describe 
whether the sum rule is valid above and below messenger scale respectively. The bottom three sum rules involving scalar masses are valid only above messenger scale and with the condition of unimodular weights $n_Q=n_U=n_D=n_E=n_L=n_u$. Above the messenger scale $b_a'$ is to be substituted for $b_a$.} 
\label{sumrule}
\end{table}

We will attempt to derive sum rules valid at all energies using the RGIs, that are not dependent on other parameters than $n_i$ and $N$. From Table \ref{coef} we see that  $D_{L_{13}}=0$ and thus we
immediately have
\begin{equation}
2(m_{\tilde{L}_1}^2-m_{\tilde{L}_3}^2)-m_{\tilde{e}_1}^2+m_{\tilde{e}_3}^2=0.
\label{sumDL}
\end{equation}

More sum rules can be constructed by combining the solutions to the mass parameters and $g$.  For instance by equating the two expressions for $\mgr$; (\ref{m32sol}), and (\ref{m32solalt}) with $a=1$, we obtain 
\begin{align}
\frac{165}{48} g^4 \gGUT^4 N (5 N+33) (\ib{1}-\ib{3})=&\sqrt{5}g^2 \gGUT^2\left(g^2 (5 N+33)-33 \gGUT^2\right)\sqrt{N\left(\imcal{1} \nbeta (5 N+33)-\frac{33}{\nbeta} \dx\right)}\rv+5 g^4 \gGUT^4 \ib{1} N (5 N+33)-165 g^2 \gGUT^4 N \sqrt{\frac{\dx}{\nbeta}},
\end{align}
where we have again used $M_0^2=\frac{\dx}{\nbeta}$ and $g^2=\frac{-\mathcal{Y}(\ndg)}{10 \dya}$. Similarly from the solutions of $g^2$; (\ref{gsol1}) and (\ref{gsolalt}), we obtain
\begin{align}
&-10n_{\delta\gamma} \dx-13 \dy n_{\beta}\rv=\frac{160 \pi^2\gGUT^2\dya\left(n_{\beta}\mathcal{I}_{M_2}-\dx\right)}{16 \pi ^2\left((1+N)\mathcal{I}_{M_2}-\frac{a\dx}{\nbeta}\right)+\mathcal{I}_{B_2} \sqrt{\frac{\dx}{\nbeta}} N-\left(\mathcal{I}_{B_2}+16 \pi ^2 \sqrt{\frac{\dx}{\nbeta}}\right)\sqrt{N \left((1+N) \mathcal{I}_{M_a}-\frac{\dx}{\nbeta}\right)}}.
 \end{align}

To construct sum rules independent on $N$, one can combine to solutions (\ref{Nsol}) to form three sum rules
\begin{align}
N_{(1)}=N_{(2)}=N_{(3)}.
\end{align}
We leave out the explicit formula for brevity. This provides a way to determine some of the modular weights. Additional sum rules can be formed in a similar manner from (\ref{m32sol1}), (\ref{m32sol2}), (\ref{M0sol1}), and (\ref{M0solalt}) or by combining other solutions summarized in Table \ref{inv3}.

\subsection{Special cases}
In order to draw distinctions between the supersymmetry breaking scenarios and identify dominating contributions we consider sum rules in some special cases obtained from the deflected mirage mediation boundary conditions.

\subsubsection{The case $N=0$}
In the absence of messenger fields, $\Delta \ib{a}=\Delta \im{a}=0$, and all the invariants have equal values at the TeV-scale and the GUT-scale. As noted before, since the threshold correction for $\ib{a}$,
\begin{align}
\Delta \ib{a}=-NM_0/(16 \pi^2)(1+\alpha_g)\alpha_m \ln \frac{M_P}{m_{3/2}},
\end{align}
vanishes for $\alpha_g=-1$ as well as for $N=0$, we cannot deduce  the absence of the messenger fields from the condition $\ib{a}(\mutev)=\ib{a}(\mugut)$ alone.
On the other hand, from \rf{deltaim1}, the threshold correction for $\im{1}$ reads
\begin{align}
\Delta \im{1}=&-2M_1\Delta M_1-\Delta M_1^2+\frac{5N}{8}(m_{\tilde{d}_1}^2(\mumes^+)-m_{\tilde{u}_1}^2(\mumes^+)-m_{\tilde{e}_1}^2(\mumes^+))\nonumber\\&-\frac{5b_1}{8}(\Delta m_{\tilde{d}_1}^2-\Delta m_{\tilde{u}_1}^2-\Delta m_{\tilde{e}_1}^2). \label{deltaim1a}
\end{align}
If $\ag=-1$, the corrections to scalar and gaugino masses vanish, leaving
\begin{align}
\Delta \im{1}=\frac{5N}{8}(m_{\tilde{d}_1}^2(\mumes^+)-m_{\tilde{u}_1}^2(\mumes^+)-m_{\tilde{e}_1}^2(\mumes^+)), \quad {\rm for} \  \ag=-1.
\end{align}
As one can see, $\Delta \im{1}=0$ only if $N=0$ (or $m_{\tilde{d}_1}^2-m_{\tilde{u}_1}^2-m_{\tilde{e}_1}^2=0$, which would imply a constant $M_1$ above the messenger scale). Similar condition obviously applies for $\im{2}$ and $\im{3}$. Consequently, the equation $\im{a}(\mugut)=\im{a}(\mutev)$ provides us with a better condition for the absence of messengers. Thus we write,
\begin{align}
\im{a}(\mutev)=\im{a}(\mugut)=M_0^2(1+b_a\nepsilon{a}), \quad \text{for } N=0. \label{conditionN0}
\end{align}
From \rf{nepsilon1}-\rf{nepsilon3} we see that in the case of universal modular weights $\nepsilon{1}=\nepsilon{2}=2\nepsilon{3}=5/8(1-n_u)$. Now the $\nepsilon{a}$ can be eliminated from \rf{conditionN0} and three conditions obtained,
\begin{align}
&M_0=\frac{\sqrt{3\im{2}+2\im{3}}}{\sqrt{5}},\\
&5/8(1-n_u)=\frac{2 (\im{2}-\im{3})}{3 \im{2}+2 \im{3}},\\
&81 \im{2}-56 \im{3}-25\im{1}=0.\label{conditionN0a}
\end{align}
The equation \rf{conditionN0a} can in fact also be derived from the assumptions of scalar mass and gaugino mass unification at the same scale without additional assumptions about the supersymmetry breaking scenario  \cite{Hetzel:2012bk}. We then use the solution $M_0^2=\frac{\dx}{5(1-n_u)}$ from \rf{M0sol}, and the first two equations yield (after eliminating $n_u$ and $M_0$),
\begin{align}
\dx=16/5(\im{2}-\im{3}),\quad\text{for } N=0, n_u\neq1. \label{conditionN0b}
\end{align}
The equations \rf{conditionN0a} and \rf{conditionN0b} now provide a simple test for determining the existence of the messenger sector, with the caveat that universal modular weights are required.

\subsubsection{The cases $M_0=0$, $\mgr=0$, $g=\gGUT$}
In the case of zero $M_0$ and $\mgr$ we obtain from \rf{M0sol}, \rf{m32sol},
\begin{align}
&\dx=0, \quad{\rm for}\ M_0=0, \label{M0sum0}\\
&\sum_{b,c=1}^3\epsilon_{abc}\ib{c}=0, \quad  (a=1,2,3), \quad {\rm for} \ \mgr=0. \label{m32sum0}
\end{align}
 The case $g=\gGUT$ implies either $\mumes=\mugut$ or $N=0$. From \rf{gsol}
 \begin{align}
13\dy+10M_0^2(n_{\gamma}+n_{\delta})=10\gGUT^2\dya.
\end{align}
If we assume universal modular weights this is simplified to
\begin{align}
13\dy=10\gGUT^2\dya. \label{gsum}
\end{align}

\section{Comparison with SUGRA, AMSB, gauge mediation and pure Mirage mediation}
\label{sec:compare}
Deflected mirage mediation includes contributions from three separate supersymmetry breaking mechanisms, namely gravity mediation (SUGRA) \cite{Chamseddine:1982jx, Barbieri:1982eh, Ibanez:1982ee, Hall:1983iz, Ohta:1982wn}, gauge mediation (GMSB) \cite{hep-ph/9303230, hep-ph/9408384, hep-ph/9507378}, and anomaly mediation (AMSB) \cite{hep-th/9801155, hep-ph/9810442}. The boundary conditions for the scalar and the gaugino masses can be parametrized as
\begin{align}
\text{SUGRA:}&\quad m_i^2(\mugut)=(1-n_i) M_0^2; \qquad\qquad\qquad\qquad M_a^2(\mugut)=M_0^2,\label{sugra}\\
\text{AMSB:}&\quad m_i^2(\mugut)=- \frac{\dot{\gamma'_i}}{(16\pi^2)^2}\mgr^2; \qquad\qquad M_a^2(\mugut)=g^2\frac{b_a}{16\pi^2}m_{3/2},\label{amsb}\\
\text{GMSB:}&\quad m_i^2(\mumes)=\frac{N\Lambda^2}{(16\pi^2)^2}\sum_{a=1}^3 c_a(\Psi_i)g_a^4; \quad M_a^2(\mumes)=-\frac{Ng_a^2}{16\pi^2}\Lambda. \label{gmsb}
\end{align}
Note that the GMSB boundary conditions are defined on the messenger scale possibly different from the GUT scale while the AMSB and SUGRA boundary conditions are defined at the GUT-scale. Additionally, 
two combinations of the above exist: mirage mediation (MMSB) \cite{Kachru:2003aw} is obtained from DMMSB by removing the messengers and deflected anomaly mediation (DAMSB)
\cite{Pomarol:1999ie, Okada:2002mv,Wang:2015nra} is obtained by setting $M_0$ to zero.

The boundary conditions for the five models  can be extracted from (\ref{gauginoUV}), (\ref{scalarUV}), (\ref{deltam}), and (\ref{deltaM}) with the 
following prescriprions:

\begin{align}
\text{SUGRA:}&\quad \mgr=0; \: N=0,\\
\text{AMSB:}&\quad  M_0=0; \: N=0,\\
\text{GMSB:}&\quad M_0=0; \: \mgr=0, \\
\text{Mirage:}&\quad N=0,\\
\text{DAMSB:}&\quad M_0=0.
\end{align}

Here we assume universal modular weights for all applicable models and the parameters not specified to be nonzero. In the case that one of the mechanisms clearly dominates supersymmetry breaking scenario can then in principle be resolved or narrowed by measuring the parameters in terms of RGIs. By comparing to the solutions for the parameters in terms of invariants (\ref{m32sol}), (\ref{M0sol}), and (\ref{tmessol}), along with the sum rules \rf{conditionN0a}-\rf{gsum}, we can deduce
from the measured invariants $\ib{1}$, $\ib{3}$, $\im{a}$, $\dya$, and $\dy$ the following:
\begin{align}
I_{B_1}-I_{B_3}\propto\mgr&\begin{cases}0, \text{ for mSUGRA, GMSB}\\>0, \text{ for AMSB,Mirage, DAMSB}\end{cases},\\
\dx=(1-n_u)M_0^2&\begin{cases}=0, \text{ for AMSB, GMSB}\\\propto(1-n_u), \text{ for mSUGRA, Pure Mirage, DAMSB}\end{cases},\\
\frac{10 \dya}{13 \dy}+\frac{1}{g_{\text{GUT}}^2}=\frac{1}{\gGUT^2}-\frac{1}{g^2}&\begin{cases}=0 \text{ for mSUGRA, AMSB, Pure Mirage}\\\propto \tmes N,{\text{ for GMSB, DAMSB}}
\end{cases},\\
81 \im{2}-56 \im{3}-25\im{1}&\bc =0,\quad\text{for AMSB, mSUGRA, Pure Mirage}\\ \neq0,\quad\text{for Deflected Mirage, GMSB, DAMSB}\ec .
\label{4inv}                         
\end{align}

Thus e.g. observing $\dx=0$ would exclude mediation mechanism with a gravity mediated contribution with $n_u\neq1$, but deflected or pure mirage with $n_u=1$ cannot be ruled out. On the other hand  a nonzero $\frac{10 \dya}{13 \dy}+\frac{1}{g_{\text{GUT}}^2}$ implicates a gauge mediated contribution, with the messenger scale different from the GUT scale. We have illustrated the implications of different values of (\ref{4inv}) in Fig. \ref{flowchart} by starting from the measurement of $\dx$.

\tikzstyle{startstop} = [rectangle, rounded corners, minimum width=3cm, minimum height=1cm, text width=3.2cm, text centered, draw=black, fill=red!30]
\tikzstyle{startstoplong} = [rectangle, rounded corners, minimum width=3cm, minimum height=1cm, text width=3.5cm, text centered, draw=black, fill=red!30]
\tikzstyle{startstopshort} = [rectangle, rounded corners, minimum width=1cm, minimum height=1cm, text width=2cm, text centered, draw=black, fill=red!30]
\tikzstyle{startstoplarge}= [rectangle, rounded corners, minimum width=1cm, minimum height=1.3cm, text width=2.5cm, text centered, draw=black, fill=red!30]
\tikzstyle{io} = [trapezium, trapezium left angle=70, trapezium right angle=110, minimum width=3cm, minimum height=3cm, text centered, draw=black, fill=blue!30]
\tikzstyle{process} = [rectangle, minimum width=3cm, minimum height=1cm, text centered, draw=black, fill=orange!30]
\tikzstyle{decision} = [diamond, minimum width=2cm, minimum height=2cm, text centered, draw=black, fill=green!30]
\tikzstyle{arrow} = [thick,->,>=stealth]
\begin{figure}
\centering
\begin{tikzpicture}[node distance=4cm]
\node (start2) [decision] {$\dx$};
\node (dg1) [decision, below of=start2, xshift=1.5cm] {$\frac{1}{\gGUT^2}-\frac{1}{g^2}$};
\node (ib1) [decision, below of=dg1] {$I_{B_3}-I_{B_1}$};
\node (dmi1) [startstop, right of=dg1, xshift=0.5cm]{Deflected mirage};
\node(amsb2)[startstop, right of=dg1, xshift=0.5cm, yshift=2.7cm]{AMSB};
\node (pmi1) [startstop, right of=ib1, xshift=0.5cm]{Pure mirage};
\node (sugra1) [startstoplong, below of=ib1, xshift=-1cm, yshift=-1cm]{$n_u$=1\:\:SUGRA\:\:\:\:\:\:\:\:\:};
\node(DAMSB1) [startstop, below of=dmi1, yshift=2.5 cm]{DAMSB};
\node(DAMSB2) [startstop, above of=amsb2, yshift=-1.7 cm]{DAMSB\\$\mugut=\mumes$};

\node (DZ) [decision, below of=start2, xshift=-1.5cm] {$I_{B_3}-I_{B_1}$};
\node (dg2) [decision, below of=DZ] {$\frac{1}{\gGUT^2}-\frac{1}{g^2}$};
\node(dmi3)[decision, left of=DZ, xshift=-1cm]{$\frac{1}{\gGUT^2}-\frac{1}{g^2}$};
\node(amsb)[startstop, left of=start2, xshift=-2.5cm, yshift=2cm]{D. Mirage with $\mumes=\mugut$};
\node(Ntest1)[decision, left of=start2, xshift=-2.5cm, yshift=-1.3cm]{$\sum f_a\im{a}$};
\node(dmi5)[startstoplarge, below of=dmi3, xshift=-0cm,, yshift=0.1cm]{Deflected Mirage\\ $\mgr$=0\\ };
\node (GMSB) [startstopshort, below of=pmi1, yshift=-0.2cm]{GMSB};
\node (dmi4) [startstop, below of=dmi5, xshift=0cm, yshift=-1cm]{D. Mirage with $\mumes=\mugut$, $\mgr=0$};
\node(Ntest2)[decision, below of=dg2, yshift=1cm]{$\sum f_a\im{a}$};
\draw [arrow] (start2) -- node[anchor=west]  {$\neq0$}(dg1);
\draw [arrow] (dg1) -- node[anchor=south] {$\neq0$} (dmi1);
\draw [arrow] (dg1) -- node[anchor=west]  {$=0$} (ib1);
\draw [arrow] (ib1) -- node[anchor=south]  {$\neq0$} (pmi1);
\draw [thick] (ib1) -- node[anchor=west] {$=0$}(1.5,-12.0);
\draw [arrow] (1.5,-12.2) --(1.5,-12.5);
\draw[thick] (Ntest2)--node[anchor=south]{$\neq0$}(-6.5,-11);
\draw[arrow](-6.5,-11)--(dmi4);
\draw[thick](Ntest1)--node[anchor=south] {$=0$}(-0.6,-1.3);
\draw[thick](-0.4,-1.3)--(0.4,-1.3);
\draw[arrow](0.6,-1.3)--(amsb2);
\draw [arrow] (start2) -- node[anchor=east]  {$=0$} (DZ);
\draw [arrow] (DZ) -- node[anchor=west]  {$=0$} (dg2);
\draw [thick] (dg2) --node[anchor=west] {$=0$}(Ntest2);
\draw[arrow](-1.4,-12.1)--(GMSB);
\draw[thick](-6.5,-12.1)--(-1.6,-12.1);
\draw[thick, dashed,red](start2)--node[anchor=east] {$n_u=1$}(0,-12.5);
\draw[thick, dashed,red](start2)--(0,2.5);
\draw(0,-12.5)--(0,-13.5);
\draw(-7.85,-7.9)--(-5.15,-7.9);
\draw[arrow, red](0,-7.5)--(-0.3,-7.5);
\draw[arrow](-2.7,-8)--node[anchor=south]  {$\neq0$}(-5.15, -8);
\draw[arrow](DZ)--node[anchor=south]{$\neq0$}(dmi3); 
\draw[arrow](dmi3)--node[anchor=west]{$\neq0$}(dmi5); 
\draw[arrow](dmi3)--node[anchor=west]{$=0$}(Ntest1); 
\draw[arrow](Ntest1)--node[anchor=west]{$\neq0$}(amsb);
\draw[thick](Ntest2)--node[anchor=east]{$=0$}(-1.5,-13);
\draw[arrow](-1.5,-13)--(sugra1);
\draw[thick](-6.5, -5.5)--(-1.6,-5.5);
\draw[thick](-1.3, -5.5)--(1.3, -5.5);
\draw[arrow](1.6, -5.5)--(DAMSB1);
\draw[arrow](-6.5,1)--(DAMSB2);
\end{tikzpicture}
\caption{ Implications of measurement of RGIs for the supersymmetry breaking mechanism with the assumption of $n_h=n_{H_d}=n_{H_u}$ and $n_u=n_U= n_D = n_L = n_E =n_Q$, assuming that one or two of the mechanisms dominate. On the left side of the red dotted line all endpoints have $n_u=1$. $\frac{1}{g^2} =-\frac{10 \dya}{13 \dy}$ and $\sum f_a\im{a}=81 \im{2}-56 \im{3}-25\im{1}$.}
\label{flowchart}
\end{figure}

\section{Conclusions}
\label{sec:summary}
Deflected mirage mediation is the most general type of mechanism for spontaneous supersymmetry breaking in the sense that it includes contributions from three SSB mechanisms, namely gravity-, anomaly-, and gauge mediation. The renormalization group invariants provide a way of determining the values of the supersymmetry breaking parameters, but in the case of DMMSB, the emergence of gauge messenger fields at a scale possibly different from the GUT-scale complicates their use by inducing corrections to the gaugino and the scalar masses and modifying the beta functions at this threshold. Thus the invariants have differing values above and below the messenger scale. In order to connect the TeV scale measurements of the particle masses to the GUT-scale parameters we have derived the threshold corrections to the RGIs and derived the RGIs for arbitrary $b_a$-coefficients of the beta functions. It is shown that the high scale parameters which include $N$, $\mumes$, $\mgr$, $
 M_0$, and $\alpha_g$ can be analytically solved in terms of the RGIs, and the explicit formulas are provided. We have then examined various limits of DMMSB to see how any of the contributing three pure supersymmery breaking scenarios can be detected by measuring the RGIs at the TeV scale.
\par We have also discussed how the solutions to the supersymmetry breaking parameters can be used to construct sum rules that would allow further testing of the theory and determine the modular weights for the scalar masses.
The lightest Higgs mass measurement allows the restriction of the parameter space of DMMSB. We have discussed the implications of the measured Higgs mass of 125.7 GeV to the mass spectrum and the parameters of DMMSB.

\bigskip
\section{Acknowledgments}
KH and PT acknowledge support from the Academy of Finland (Project No 137960).
PNP would like to thank Department of Physics, University of Helsinki, and
Helsinki Institute of Physics for hospitality
while part of this work was done. The work of PNP is supported by
Raja Ramanna Fellowship of the Department of Atomic Energy, and partly by  
J. C. Bose National Fellowship of the Department of 
Science and Technology, India.


\end{document}